\documentclass[twocolumn]{aastex63}
\usepackage{longtable,amsmath}
\usepackage[T1]{fontenc}

\newcommand{\kms}{km~s$^{-1}$}
\newcommand{\Msun}{M$_\sun$}

\newcommand{\Ha}{H$\alpha$}
\newcommand{\Hb}{H$\beta$}
\newcommand{\Hg}{H$\gamma$}
\newcommand{\swift}{\textsl{Swift}}
\newcommand{\HeII}{\ifmmode {\rm He}\,{\sc ii}\,\lambda4686
    \else He\,{\sc ii}\,$\lambda4686$\fi}

\accepted{August 13 2020}
\submitjournal{ApJ}
\shortauthors{Hung et al.}
\graphicspath{{./}{figures/}}

\begin{document}

\title{Double-Peaked Balmer Emission Indicating Prompt Accretion Disk Formation in an X-Ray Faint Tidal Disruption Event}

\correspondingauthor{Tiara Hung}
\email{tiarahung@ucsc.edu}

\author[0000-0002-9878-7889]{Tiara Hung}
\affiliation{Department of Astronomy and Astrophysics,
University of California, Santa Cruz, California, 95064, USA}

\author{Ryan~J.~Foley}
\affiliation{Department of Astronomy and Astrophysics,
University of California, Santa Cruz, California, 95064, USA}

\author{Enrico~Ramirez-Ruiz}
\affiliation{Department of Astronomy and Astrophysics,
University of California, Santa Cruz, California, 95064, USA}
\affiliation{DARK, Niels Bohr Institute, University of Copenhagen, Lyngbyvej 2, 2100 Copenhagen, Denmark}

\author{Jane~L.~Dai}
\affiliation{Department of Physics, University of Hong Kong}
\affiliation{DARK, Niels Bohr Institute, University of Copenhagen, Lyngbyvej 2, 2100 Copenhagen, Denmark}

\author{Katie~Auchettl}
\affiliation{School of Physics, The University of Melbourne, Parkville, VIC 3010, Australia}
\affiliation{ARC Centre of Excellence for All Sky Astrophysics in 3 Dimensions (ASTRO 3D), Australia}
\affiliation{Department of Astronomy and Astrophysics,
University of California, Santa Cruz, California, 95064, USA}
\affiliation{DARK, Niels Bohr Institute, University of Copenhagen, Lyngbyvej 2, 2100 Copenhagen, Denmark}

\author{Charles~D.~Kilpatrick}
\affiliation{Department of Astronomy and Astrophysics,
University of California, Santa Cruz, California, 95064, USA}

\author{Brenna~Mockler}
\affiliation{Department of Astronomy and Astrophysics,
University of California, Santa Cruz, California, 95064, USA}

\author{Jonathan~S.~Brown}
\affiliation{Department of Astronomy and Astrophysics,
University of California, Santa Cruz, California, 95064, USA}

\author{David~A.~Coulter}
\affiliation{Department of Astronomy and Astrophysics,
University of California, Santa Cruz, California, 95064, USA}

\author{Georgios~Dimitriadis}
\affiliation{Department of Astronomy and Astrophysics,
University of California, Santa Cruz, California, 95064, USA}

\author{Thomas~W.-S.~Holoien}
\affiliation{The Observatories of the Carnegie Institution for Science, 813 Santa Barbara Street, Pasadena, CA 91101, USA}

\author[0000-0001-8825-4790]{Jamie~A.P.~Law-Smith}
\affiliation{Department of Astronomy and Astrophysics,
University of California, Santa Cruz, California, 95064, USA}

\author{Anthony~L.~Piro}
\affiliation{The Observatories of the Carnegie Institution for Science, 813 Santa Barbara Street, Pasadena, CA 91101, USA}

\author{Armin~Rest}
\affiliation{Space Telescope Science Institute, Baltimore, MD 21218, USA}

\author{C\'{e}sar~Rojas-Bravo}
\affiliation{Department of Astronomy and Astrophysics,
University of California, Santa Cruz, California, 95064, USA}

\author{Matthew~R.~Siebert}
\affiliation{Department of Astronomy and Astrophysics,
University of California, Santa Cruz, California, 95064, USA}

\begin{abstract}

We present the multi-wavelength analysis of the tidal disruption event (TDE) AT~2018hyz (ASASSN-18zj). From follow-up optical spectroscopy, we detect the first unambiguous case of resolved double-peaked Balmer emission in a TDE. The distinct line profile can be well-modelled by a low eccentricity ($e\approx0.1$) accretion disk extending out to $\sim$100 $R_{p}$ and a Gaussian component originating from non-disk clouds, though a bipolar outflow origin cannot be completely ruled out. Our analysis indicates that in AT~2018hyz, disk formation took place promptly after the most-bound debris returned to pericenter, which we estimate to be roughly tens of days before the first detection. Redistribution of angular momentum and mass transport, possibly through shocks, must occur on the observed timescale of about a month to create the large \Ha-emitting disk that comprises $\lesssim$5\% of the initial stellar mass. With these new insights from AT~2018hyz, we infer that circularization is efficient in at least some, if not all optically-bright, X-ray faint TDEs. In these efficiently circularized TDEs, the detection of double-peaked emission depends on the disk inclination angle and the relative strength of the disk contribution to the non-disk component, possibly explaining the diversity seen in the current sample.

\end{abstract}


\section{Introduction}
\label{sec:intro}

When a hapless star approaches too close to a supermassive black hole (SMBH) it will be torn apart by tidal forces \citep{Hills1975,Rees1988}. After the disruption, a nascent accretion disk is expected to form and produce a luminous flare \citep{Hills1975,1976MNRAS.176..633F}. These bright transients have been observed at the centers of quiescent galaxies, and evolve distinctly from supernovae, leading to their identification as tidal disruption events (TDEs).

In the past decade, wide-field ground-based optical time domain surveys have discovered tens of TDEs, which affords us the opportunity to study these rare phenomena extensively across the electromagnetic spectrum. Until now, only a small fraction ($\sim$10\%) of TDEs is detected with X-ray emission, where the presence of an accretion disk can be confidently established \citep[e.g.][]{2015Natur.526..542M,Holoien2016,2018MNRAS.474.3593K}. For the majority of the optically discovered TDEs, there has been no clear proof \citep{Gezari2012,2016MNRAS.463.3813H,2017ApJ...842...29H} that the infalling stellar debris are able to circularize and form an accretion disk \citep{2015ApJ...804...85S,2016MNRAS.455.2253B,Gezari2017a,2014ApJ...783...23G,2015ApJ...809..166G,Dai13, Dai15}. Whether an accretion disk can form promptly following a TDE has long been a debate as alternative mechanisms such as stream-stream collision are thought to be capable of powering the UV/optical emission in an optically discovered TDE even when disk formation is inefficient \citep{2015ApJ...806..164P,2017MNRAS.464.2816B}.

Here we report the tell-tale evidence of an accretion disk in a nearby TDE called AT~2018hyz (also known as ASASSN-18zj).
AT~2018hyz is a TDE first detected on 6 Nov 2018 with an apparent $V$-band magnitude of 16.4 by the All-Sky Automated Survey for Supernovae \citep{2014ApJ...788...48S}. The transient
aligns with the nucleus of the galaxy 2MASS J10065085+0141342 at $z=0.0457$ that is abscent of pre-flare nuclear activity. An archival Sloan Digital Sky Survey (SDSS) spectrum of the host galaxy displays strong Balmer absorption lines that are characteristic of an E+A galaxy, a rare subtype of post-starburst galaxies in which TDEs are preferentially discovered \citep{Arcavi2014,2016ApJ...818L..21F,2017ApJ...850...22L,2018ApJ...853...39G}. Following the spectroscopic classification of AT~2018hyz as a TDE \citep{2018ATel12198....1D}, we triggered photometric and spectroscopic monitoring spanning about a year in time.

The discovery of double-peaked emission features in the spectra of AT~2018hyz is a strong indication of an elliptical accretion disk \citep{1989ApJ...339..742C,1995ApJ...438..610E} orbiting the SMBH, though alternative scenarios such as a bipolar outflow are not entirely ruled out. Our findings suggest that the infalling debris in TDEs begin forming an accretion disk soon after the most highly bound material falls back.
This accretion powers the TDE, which is a sign of the presence of an otherwise dormant SMBH and a powerful diagnostic of its properties.

In this paper, we present new insights from the first TDE with well-separated, double-peaked emission line profile with high S/N. This paper is structured as follows, we describe the follow-up photometric data obtained by \swift\ and the Swope Telescope, and optical spectra in \autoref{sec:obs}. We detail and present the results of emission line modelling in \autoref{sec:results}. Our discussions and conclusions are presented in \autoref{sec:discussion} and \autoref{sec:conclusion}.

\section{Observations and Data Reduction}
\label{sec:obs}

Throughout the paper, we adopt a flat $\Lambda$CDM cosmology with $H_0 = 69.3$~\kms~Mpc$^{-1}$, $\Omega_m = 0.29$,
and $\Omega_\Lambda = 0.71$ measured by the Wilkinson Microwave Anisotropy Probe \citep{2013ApJS..208...20B}. The time difference ($\Delta t$) is expressed in rest-frame time with respect to the first \swift\ observation, which is close to the peak of the light curve, at MJD 58428.
All the data for AT~2018hyz presented here have been corrected for Milky Way foreground extinction assuming a Cardelli extinction curve \citep{1989ApJ...345..245C} with $R_\mathrm{V} = 3.1$ and $E(B-V) = 0.0288 \pm 0.0007$~mag \citep{2011ApJ...737..103S}.

\subsection{Photometry}
Following the spectroscopic classification of AT~2018hyz as a TDE \citep{2018ATel12198....1D}, we triggered photometric and spectroscopic monitoring spanning about 1 year in time (between November 2018 and November 2019). \autoref{fig:lightcurve} displays the light curves of AT~2018hyz observed by the Ultraviolet Optical Telescope \citep[UVOT;][]{gcg+04,2005SSRv..120...95R} and simultaneously by the X-ray Telescope (XRT) onboard the Neil Gehrels \textsl{Swift} Observatory as well as the Swope telescope at Las Campanas Observatory. We present the values of these photometry data in \autoref{tab:xmm} and \autoref{tab: phot}. The reduction of data obtained by each instrument is detailed in the following subsections.

\subsubsection{Swope Photometry}
Optical photometry of AT~2018hyz in $gri$ was obtained with the 1-meter Swope telescope from 16 Nov 2018 to 16 Jun 2019 with a 2--5-day cadence. The images were reduced using the {\texttt photpipe} imaging and photometry pipeline \citep{Rest2005, Rest2014}. We subtracted the bias and flattened each frame using bias and sky flat images obtained on the same night and in the same instrumental configuration as each AT~2018hyz image. The images were registered and geometric distortion was removed using 2MASS astrometric standards \citep{2MASS}. We measured a S/N-weighted offset of 0.5\arcsec$\pm$0.3\arcsec\ from all of our Swope images in $gri$ bands, confirming that the transient is nuclear as expected for a TDE. Using {\tt hotpants} \citep{HOTPANTS}, we subtracted pre-discovery Pan-STARRS1 (PS1) template images \citep{Flewelling16} from each Swope $gri$ frame.  For the $u$-band images, we used SDSS template images \citep{SDSSDR8}. We then obtained photometry of AT~2018hyz using a custom-built version of {\tt DoPhot} \citep{Schechter93} and a fixed point-spread function (PSF) in the difference images themselves. The photometry is calibrated using PS1 sources in the same field and transformed into the Swope natural system \citep{Scolnic15}. The final difference-image photometry of AT~2018hyz is displayed in \autoref{fig:lightcurve}.

\subsubsection{UVOT Photometry}
\begin{figure*}
\centering
\includegraphics[width=7in, angle=0]{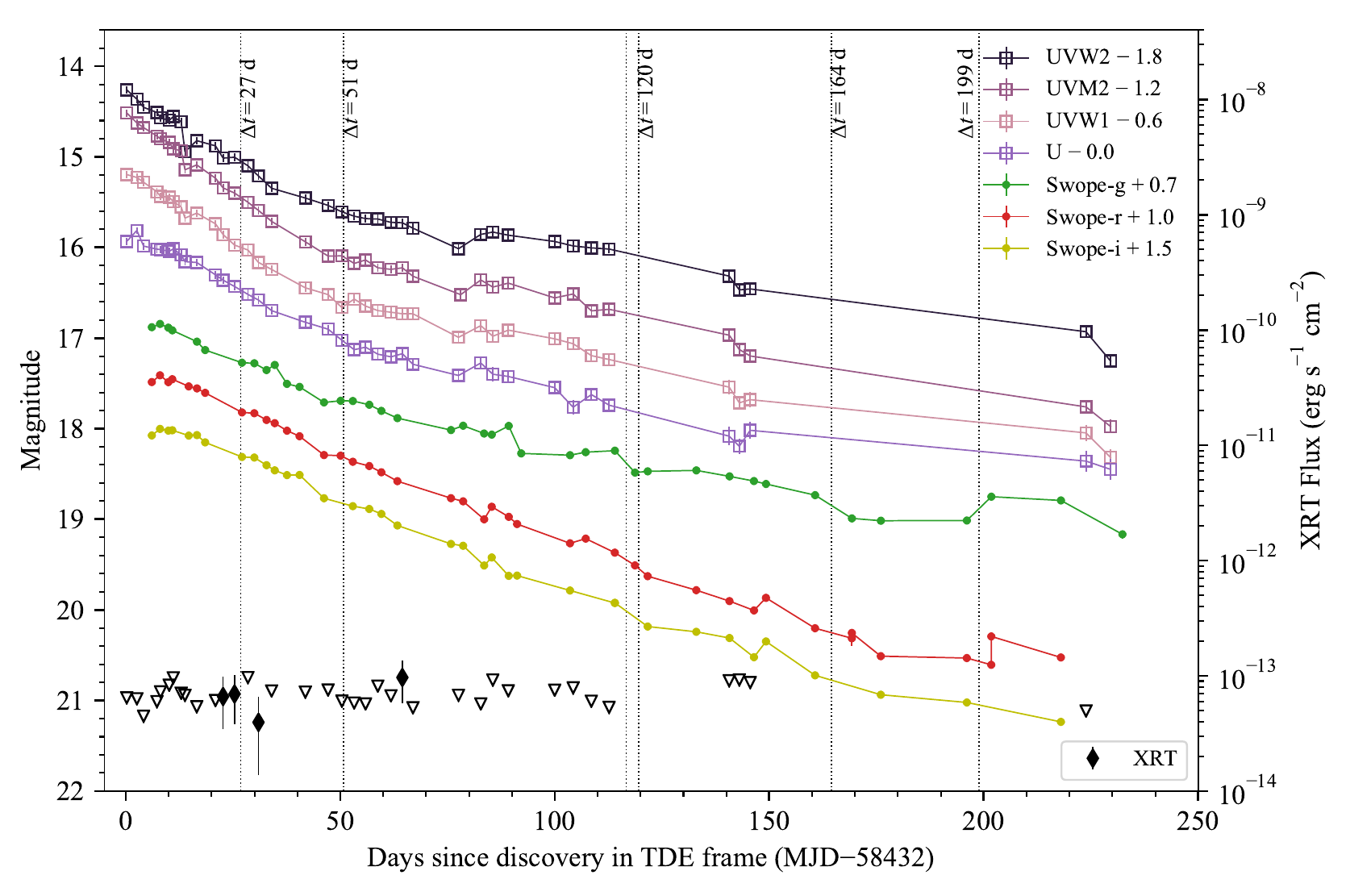}
\caption{Multi-wavelength light curves of AT~2018hyz. The host-galaxy flux has been subtracted from the Swope light curves but not the \textsl{Swift} light curves; however, the contamination in the \textsl{Swift} bands are $<$5\% of the flux.
XRT (0.3--10~keV) detections are shown as black solid diamonds with values corresponding to the vertical axis on the the right-hand side. Open black triangles mark the XRT upper limits that corresponds to $3\sigma$. The vertical dotted lines mark the epochs with spectroscopic observations. A late-time spectrum obtained on $\Delta t=364$~d is outside of our photometric coverage thus not indicated in this figure.
  \label{fig:lightcurve}}
\end{figure*}

We extracted UV light curves from a series of 38 observations obtained with the \textsl{Swift} Ultraviolet/Optical Telescope (UVOT) using a 5\arcsec\ circular aperture. Although the observations were made in all 6 UVOT filters (UVW2, UVM2, UVW1, $U$, $B$, and $V$), we do not include the \textsl{Swift} $B$- and $V$-band data in \autoref{fig:lightcurve} since galaxy light is non-negligible in these bands and host templates cannot be obtained yet. Given that the host galaxy is of an early type that contributes very little of the observed UV light ($u>19.9$ mag), we did not attempt host galaxy subtraction for the \swift\ light curves presented in \autoref{fig:lightcurve}.

\subsubsection{XRT Photometry}
Simultaneous \textsl{Swift} X-Ray Telescope (XRT) observations of AT~2018hyz were also obtained, complementing our \textsl{Swift} UVOT observations. All level-one XRT data were analyzed and reduced using the standard filters and screening criteria as suggested in the \textit{Swift} XRT data reduction guide\footnote{\url{http://swift.gsfc.nasa.gov/analysis/xrt_swguide_v1_2.pdf}} and using the \textsl{Swift} XRTPIPELINE version 0.13.2 with the most up-to-date calibration files.  To quantify the presence of X-ray emission at the position of AT~2018hyz, we used a source region with a radius of 50\arcsec centered on the position of AT~2018hyz and a source-free background region with a radius of 150\arcsec\ centered at R.A.$=$10h06m59.6s, Decl.$=+01^{\circ}47\arcmin56.3\arcsec$ (J2000). Since only a fraction of possible AT~2018hyz photons can be captured by the size of the source region, all extracted count rates are corrected for the encircled energy fraction\footnote{50\arcsec\ radius corresponds to the $\sim$90\% encircled energy radius at 1.5~keV assuming an on-axis pointing \citep{2004SPIE.5165..232M}}.

In most \textsl{Swift} XRT epochs, we do not detect X-ray emission from AT~2018hyz. In these cases, we derive a 3-$\sigma$ upper limit to the count rate. However, approximately 27, 29, 35 and 70 days after discovery we detect faint X-ray emission arising from the source.

To convert our count rate (both upper limits and detections) into a flux, we assume an absorbed power-law model with a photon index of $\Gamma=2.7$, which is the best-fitting value for the spectrum extracted from all \textsl{Swift} XRT observations combined. To merge our observations, we used XSELECT version 12.9.1c, while we used the \textsl{Swift} task XRTPRODUCTS to extract both source and background spectra. Ancillary response files were created using XRTMKARF, while we used the ready-made response matrix files that are available in the \textsl{Swift} CALDB. This merged spectrum was grouped using the FTOOLS command \textit{grppha} and assumed to have a minimum of 10 counts per energy bin. To model the spectrum we used XSPEC 12.10.1f, $\chi^2$ statistics, and assume a redshifted absorbed power law. Here we assume a column density of $2.59 \times 10^{20}$~cm$^{-2}$, which is the Galactic H\textsc{i} column density in the direction of AT~2018hyz \citep{2016A&A...594A.116H}.\\

\subsection{Optical Spectroscopy}
\begin{figure}
\begin{center}
\includegraphics[width=3.5in]{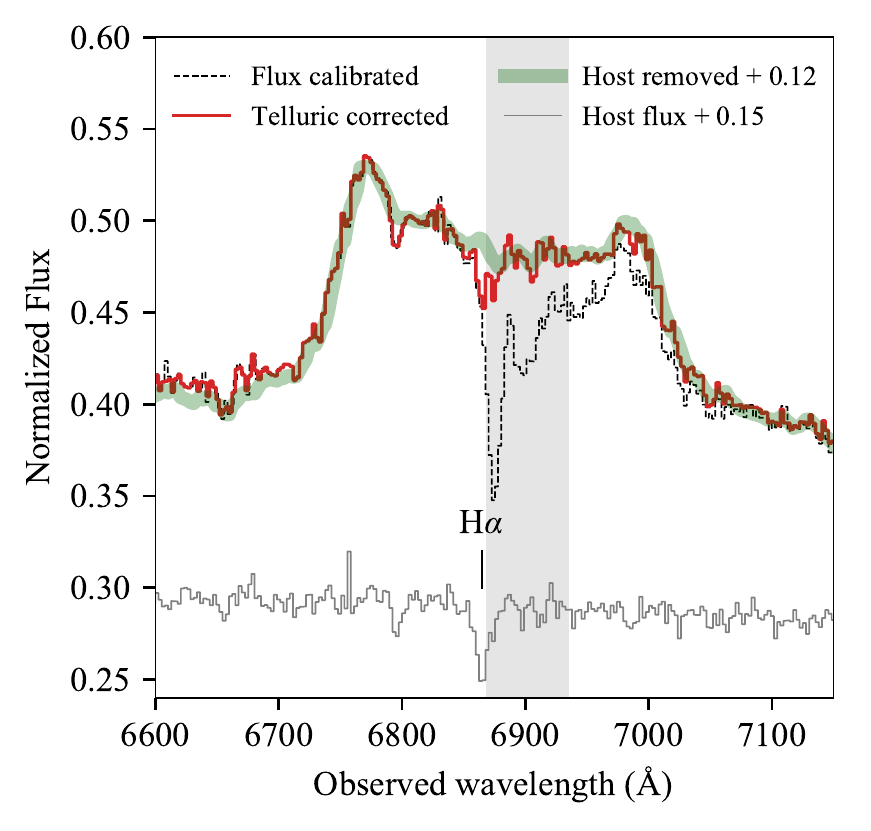}
\end{center}
\vskip -0.2in
\caption{The \Ha\ profile on $\Delta t=51$ before (black dashed line) and after (red curve) telluric absorption. The host spectrum from SDSS is plotted as thin grey line. The final host-subtracted spectrum smoothed by a boxcar of 5\AA\ used in the fitting (see \autoref{sec: e_disk_model_fit}) is shown as the thick green line. The host spectrum and the host-subtracted spectrum have been offset by a constant as indicated in the legend. The grey band marks the region that is affected by the B-band telluric absorption.
  \label{fig:telluric}}
\end{figure}

\begin{figure*}
\begin{center}
\includegraphics[width=5.8in]{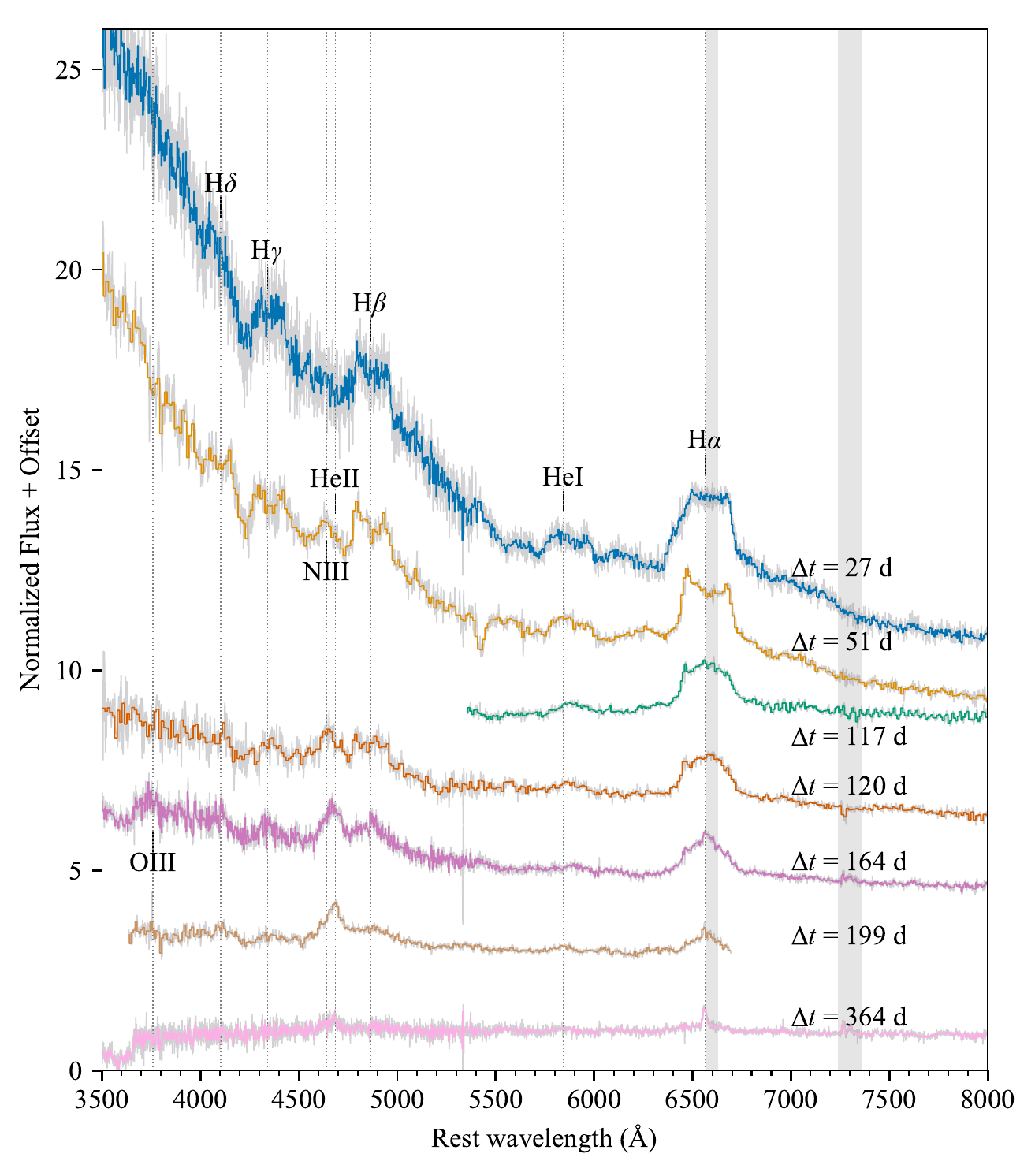}
\end{center}
\vskip -0.2in
\caption{Observed spectroscopic sequence of AT~2018hyz with host-galaxy light subtracted. The grey curves are the observed spectra while the colored solid curves are the observed spectra smoothed by 4.5~\AA. TDE emission lines are indicated by vertical grey dotted lines with labels marked. The grey shaded regions indicate the location of telluric absorption features, which have been corrected using standard-star observations. As AT~2018hyz evolves, the \ion{He}{1} emission diminishes while the Bowen emission lines (\ion{N}{3} and \ion{He}{2}) become stronger. The \Ha\ emission line evolves from a flat-topped shape in the first spectrum to double-peaked in the second to single peaked in the last spectrum on a timescale of months is unseen in other TDEs. The strong absorption feature on day 51 at around 5430~\AA\ is likely the \ion{Na}{1}~D blueshifted by $\sim$0.08c.
  \label{fig:opt_spectra}}
\end{figure*}

We obtained a total of 7 spectroscopic observations with the Kast spectrograph \citep{Miller1993} on the Lick Shane telescope, the Goodman spectrograph on the SOAR Telescope \citep{2004SPIE.5492..331C}, and the Low-Resolution Imaging Spectrometer (LRIS) \citep{1995PASP..107..375O} on the Keck~I telescope.
Detailed instrumental configurations are listed in \autoref{tab:obs_spec}. We performed standard spectrum extraction and flux absorption with standard \texttt{PyRAF}\footnote{\url{http://www.stsci.edu/institute/software_hardware/pyraf}} routines. Observations of standard stars Feige~34 and BD+284211 were used to determine the relative flux calibration and remove telluric features \citep{foley03, silverman12, dimitriadis19}. Examining the $\Delta t=51$ days spectrum in detail (\autoref{fig:telluric}), we see that the B-band telluric absorption that partially overlaps the \Ha\ emission profile is effectively removed through our data reduction process and does not introduce artificial structures in the line shape. All of the spectra have been corrected for Galactic extinction.
We calibrated each spectrum's absolute flux by comparing the $g$-band synthetic photometry of each spectrum to the photometry from Swope imaging data (including host contribution), interpolated to each spectroscopic epoch. We then subtracted the host-galaxy light using the archival SDSS spectrum after accounting for instrumental broadening. The galaxy-subtracted spectra are displayed in \autoref{fig:opt_spectra}.

\begin{table*}
\centering
\caption{Observing details of the optical spectra of AT~2018hyz}
\label{tab:obs_spec}
\bigskip
\begin{tabular}{lccccc}
\hline
\hline
Obs Date & Phase (days) & Telescope + Instrument & Slit Width & Grism/Grating & Exp Time (s) \\
\hline
      2018-12-08 &  27  &  Keck~I + LRIS &  1.0\arcsec\  &   600/4000 + 400/8500  &   230  \\
      2019-01-02 &  51  &  Shane + Kast &  2.0\arcsec\  &   452/3306 + 300/7500  &   1800 (blue) 1755 (red) \\
      2019-03-12 &  117 &  SOAR + Goodman &  1.0\arcsec\  &   400 m2  &   1500 \\
      2019-03-15 &  120 &  Shane + Kast &  2.0\arcsec\  &   452/3306 + 300/7500  &   2460 (blue) 2400 (red) \\
      2019-05-01 &  165 &  Keck~I + LRIS &  1.0\arcsec\  &   600/4000 + 400/8500  &   1500 (blue) 1396 (red) \\
      2019-06-06 &  199 &  SOAR + Goodman &  1.0\arcsec\  &   400 m1  &   1800 \\
      2019-11-26 &  364 &  Keck~I + LRIS &  1.0\arcsec\  &   600/4000 + 400/8500  &   1500 (blue) 1400 (red) \\
\hline
\hline
\end{tabular}
\end{table*}

\section{Analysis and Results}
\label{sec:results}
\subsection{Black hole Mass Estimation with MOSFiT}
\label{sec: bh mass}

We estimated the black hole mass using both theoretical modelling and the empirical $M_{\rm bh}-\sigma_*$ relation. We used the tidal disruption model implemented in \texttt{MOSFit} \citep{2018ApJS..236....6G,Mockler2019}, which assumes that the bolometric luminosity from the flare approximately follows the mass fallback rates from hydrodynamical simulations and translates it into bolometric flux assuming a constant efficiency parameter (the fallback rates used are from  \citealt{Guillochon2013}). While we do not yet have simulations that show the formation and evolution of TDE disks from realistic debris streams and their resultant light curves, this simple approximation works well to model observed TDE light curves. The model also allows for the possibility that part of the luminosity is originating from stream collisions. After converting the fallback rate to luminosity, the model takes the bolometric flux and reprocesses it using a blackbody photosphere to match our observed flux. The model assumes the photosphere evolves as a power law of the mass fallback rate. This requires the photosphere to grow with the mass fallback rate, but allows significant freedom given the range in allowed power law exponents and constants. In \cite{2016ApJ...830..125J}, the photosphere radius in their simulations of TDE outflows was fit well by a similar power-law relationship ($R_{\rm phot} \propto L^\xi$). As shown in \autoref{fig:fwhm}, the \texttt{MOSFiT} photosphere matches very well with the photosphere calculated from blackbody measurements that are agnostic to the \texttt{MOSFiT} TDE model.

Possible time delay in the onset of accretion due to either inefficient circularization or the viscous processes in the disk is accounted for in this model using a {\it viscous time} ($T_{viscous}$) parameter, which we derive to be smaller than the peak timescale.
We simultaneously fit all extinction corrected UV and host-subtracted optical photometry in \autoref{fig:lightcurve} with \texttt{MOSFiT} while prohibiting the peak luminosity to exceed the Eddington limit $L_{\rm Edd}$. This constraint is in agreement with the observations, where most of the TDEs seem to be Eddington-capped \citep{2017ApJ...842...29H,2017MNRAS.471.1694W,2017ApJ...843..106B} at the peak of their light curves. The \texttt{MOSFiT} TDE model has 8 parameters, listed in Table 1 of \cite{Mockler2019}. The best-fit values from the \texttt{MOSFiT} run for this event are presented in \autoref{tab:mosfit}.

\begin{deluxetable}{lcc}
\centering
\tablecaption{Best-fit parameters from \texttt{MOSFiT}\label{tab:mosfit}}
\tablehead{\colhead{Parameter$^a$} & \colhead{Value} & \colhead{Sys. Error$^b$}}
\startdata
$\rm t_{first \; fallback}$ (days)$^c$   & $-43_{-9}^{+8}$            & 15 \\
$\rm log_{10} R_{ph0}$               & $0.70_{-0.03}^{+0.03}$     & 0.4 \\
$\rm log_{10} T_{viscous}$ (days)    & $-1.2_{-1.3}^{+1.1}$       & 0.1 \\
$l_{\rm photo}$                      & $0.65_{-0.03}^{+0.04}$     & 0.2 \\
$\beta$                              & $1.7_{-0.2}^{+0.1}$        & 0.35 \\
$\rm log_{10} M_{\rm h} (M_{\odot})$ & $6.55_{-0.13}^{+0.17}$        & 0.2 \\
$\rm log_{10} \epsilon$      & $-2.49_{-0.05}^{+0.14}$     & 0.68 \\
$M_{\ast} (M_{\odot})$               & $1.0_{-0.1}^{+0.1}$        & 0.66 \\
\enddata
\tablecomments{$^a$ See \cite{Mockler2019} for detailed parameter definition. \\$^b$The systematic errors for this model were calculated by allowing the stellar mass-radius relation in the model to vary over a wide range of ages and metallicity combinations
\citep[using MIST stellar models - ][]{2011ApJS..192....3P,2013ApJS..208....4P,2015ApJS..220...15P,2016ApJ...823..102C,2016ApJS..222....8D}. Allowing this relation to vary approximates the changes in stellar structure expected for stars of different ages and metallicities, which is likely the largest source of systematic error in the \texttt{MOSFiT} TDE model \citep[see][]{Mockler2019}. Other potential sources of error not currently included in the error estimates include the spin of the star and the black hole, and very deep impact parameters. The magnitude of the effects of stellar or black hole spin will be less than the effects of the uncertainty in the stellar structure unless the star is spinning close to its breakup speed \citep[$>0.2\Omega_{\rm breakup}$][]{2019ApJ...872..163G}, or the impact parameter is very high \citep[$\beta > 6$][]{2019MNRAS.487.4790G}. Most stars will not reach spins of these speeds (Bouvier 2013), and the number of disruptions at a given impact parameter is expected to scale approximately with $\beta^{-1}$ \citep{Rees1988}, making disruptions at high impact parameters uncommon.\\
$^c$ Days relative to the first \swift\ detection (MJD$=58432$).}
\end{deluxetable}

\begin{figure}
\begin{center}
\includegraphics[width=3.5in]{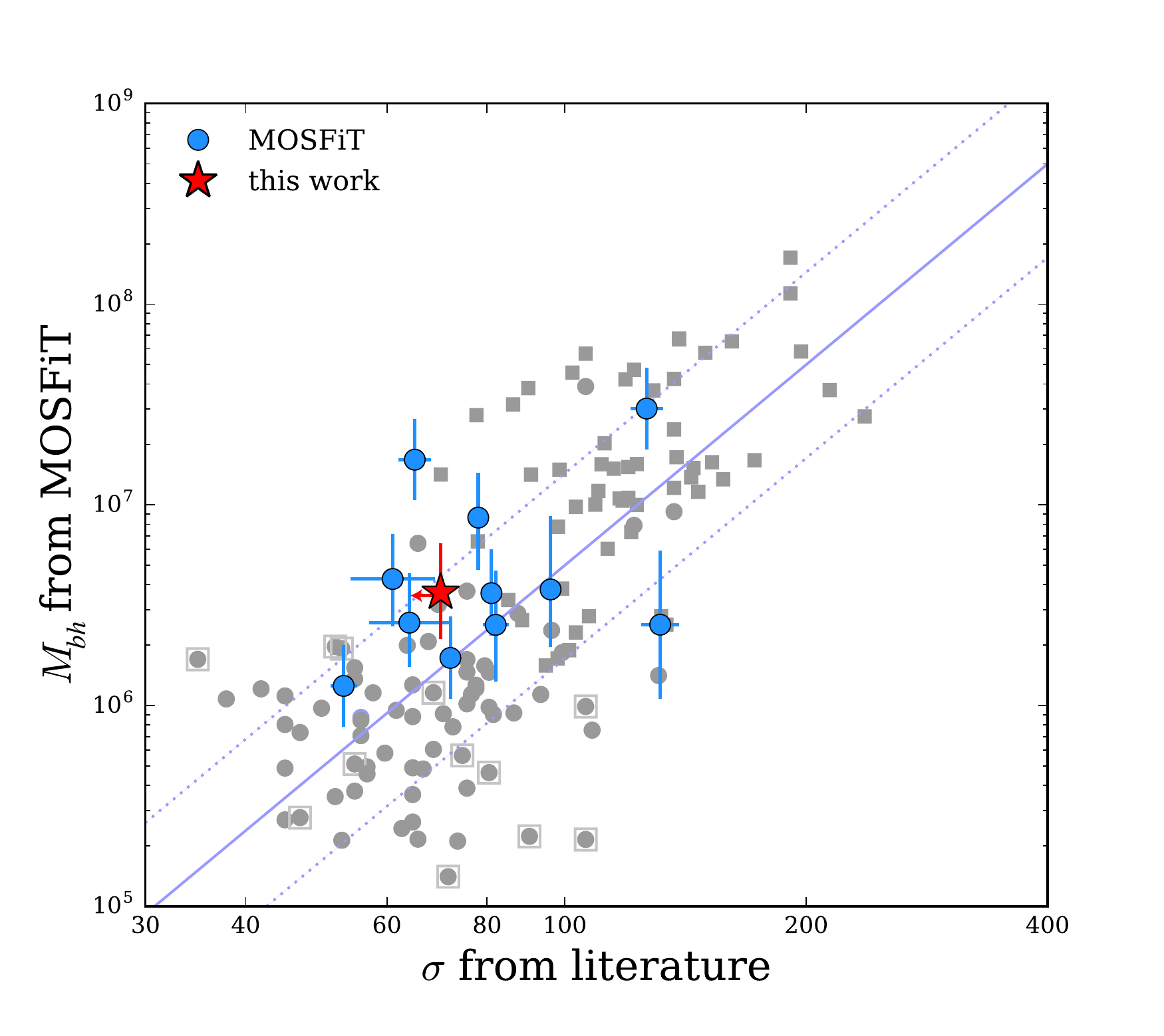}
\end{center}
\vskip -0.2in
\caption{$M_{\rm bh}-\sigma_\ast$ for host galaxies of TDEs (blue circles) and galaxies with black hole masses measured by \citet{2011ApJ...739...28X} (grey symbols; see shape notation in their paper). The derived black hole mass from this work is shown by the red star. Black hole masses estimated from \texttt{MOSFit} for TDEs are adopted from \citet{Mockler2019} and \citet{2019ApJ...887..218L}, which include a compilation of velocity dispersion across references of individual objects. The solid line marks the $M_{\rm bh}-\sigma_\ast$ in \citet{2011ApJ...739...28X}, where the dotted lines show the intrinsic scatter of 0.46 dex.
  \label{fig:msigma}}
\end{figure}

We derived a black hole mass of $\log_{10} (M_{\rm bh}/{\rm M}_{\sun})=6.55^{+0.17}_{-0.13}$ with a systematic error of 0.2~dex. In this luminosity range, \texttt{MOSFiT} determines the light-curve peak to occur at MJD 58424$^{+6}_{-4}$, about 36~days after the most-bound debris falls back.
The stellar velocity dispersion measured from the SDSS spectrum is below the SDSS spectral resolution of 70~\kms. According to the $M_{\rm bh}-\sigma_\ast$ \citep{2013ApJ...764..184M} relation, this corresponds to an upper limit of $10^{5.7}$\Msun\ on the black hole mass, with an intrinsic scatter of $\sim$0.4~dex and a measurement error of 0.2~dex. Such a small black hole mass would indicate a peak luminosity of $\approx 4 L_{\rm Edd}$ that violates our Eddington-limited constraint in \texttt{MOSFiT}. However, if we use the \citet{2011ApJ...739...28X} $M_{\rm bh}-\sigma_\ast$ relation, which was derived with a larger fraction of low-mass black holes ($M_{\rm bh}<10^7$ \Msun) than the \citet{2013ApJ...764..184M} relation (\autoref{fig:msigma}), we find a black hole mass of $10^{6.2}$\Msun\ with an intrinsic scatter of $\sim$0.5~dex, which is in better agreement with the best-fit $M_{\rm bh}$ estimated by \texttt{MOSFiT}.
It is also noteworthy that the empirical $M_{\rm bh}-\sigma_\ast$ relation is not well-constrained for the low-mass end. High-resolution spectroscopy of the host galaxy after AT~2018hyz has faded is required for a more precise determination of the stellar dispersion and hence the black-hole mass. For consistency when comparing with previously found Eddington-capped events, we adopt the black-hole mass of $3.5\times10^6$~\Msun\ from \texttt{MOSFit} throughout the paper. We also adopt this value when converting distance measurements in units of gravitational radius in our analysis.

\subsection{Temperature, Photospheric Radius, and Luminosity}
The light curves of AT~2018hyz are displayed in \autoref{fig:lightcurve}. Analysis of the \textsl{Swift} XRT observations revealed marginal soft X-ray emission (0.3--10~keV) with $L_{X}=2.8\times10^{41}$~erg~s$^{-1}$, which is almost three orders of magnitude fainter than the UV-optical emission.

We measured the temperature and the luminosity by assuming that the UV and optical emission is characterized by a blackbody spectrum on the epochs of \swift\ observations. For the first three \swift\ epochs (MJD$<58439$) that are out of the temporal coverage of Swope, we constructed the spectral energy distribution (SED) using only the $UVW2$, $UVW1$, $UVM2$, and $U$ filters. For the later \swift\ epochs, we included the optical ($g$, $r$ and $i$) photometry in the SEDs by interpolating the Swope light curves to these epochs. By modelling the SEDs with a single temperature blackbody, we find that the light curve is well described by a single-temperature blackbody ($T_{bb} = 18,000\pm2,000$~K) whose temperature evolves only mildly with time.

\begin{figure*}
\begin{center}
\includegraphics[width=3.2in]{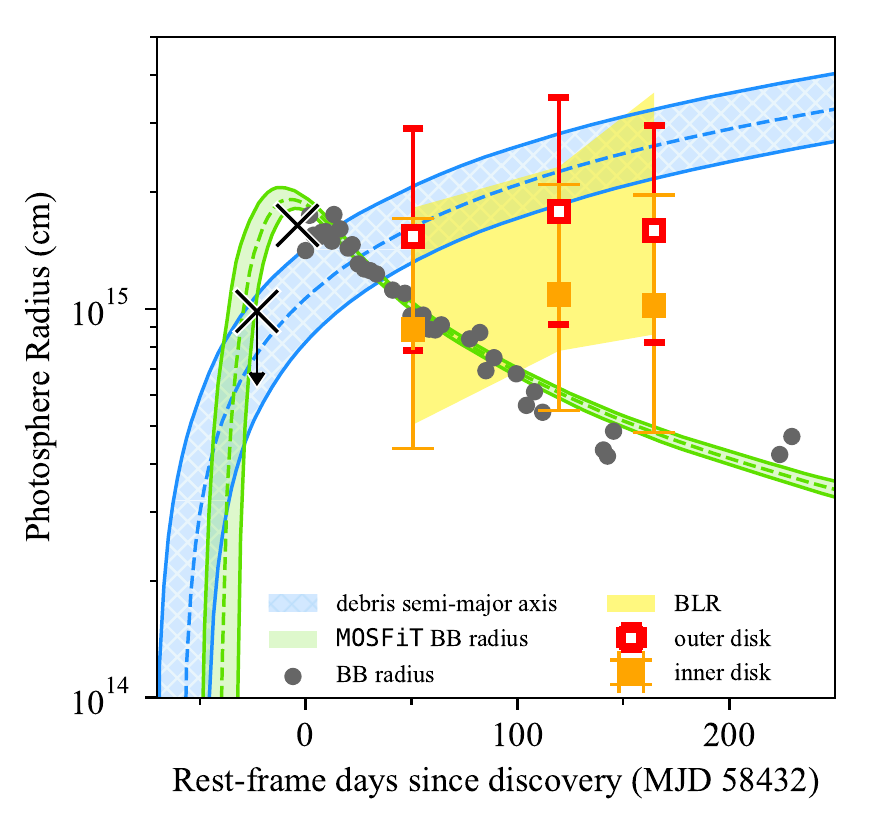}
\includegraphics[width=3.2in]{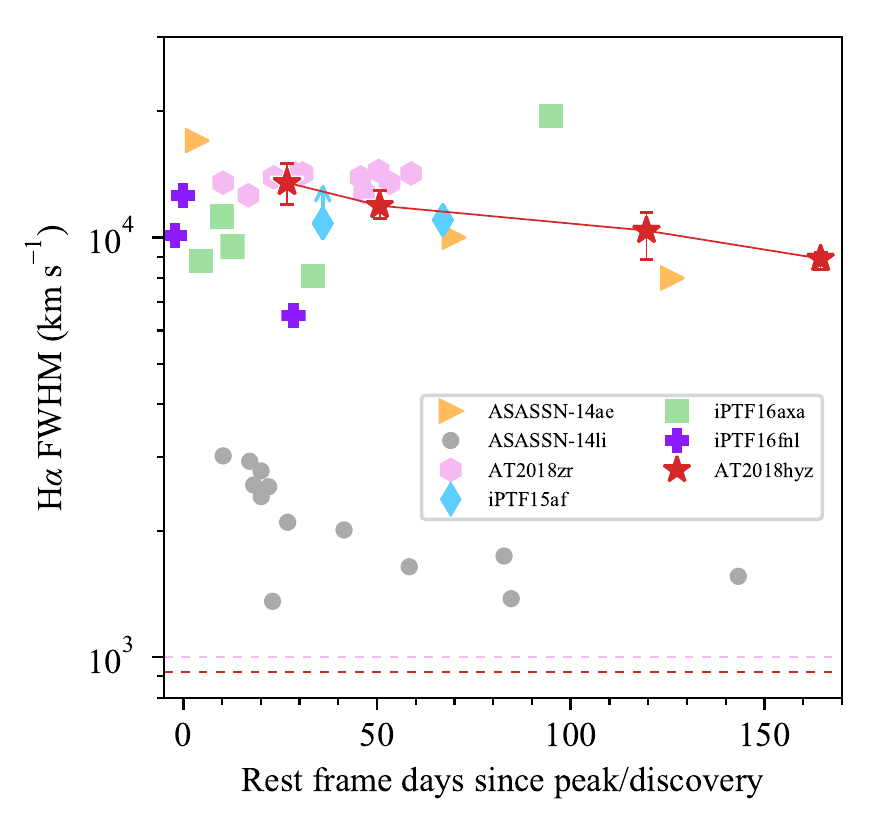}
\end{center}
\vskip -0.2in
\caption{Left: Evolution of blackbody radius for AT~2018hyz. The black circles mark the photospheric radius estimated from \textsl{Swift} photometry. The orange and red squares indicate the size of the semi-major axis of the inner and outer part of the best-fitting elliptical disk to our spectral observations, where the error bars mark the $1\sigma$ error. The broad-line region, which corresponds to the velocity width ($\sigma_{\rm g}$) of the Gaussian component seen in our spectra, is shaded in yellow. The blue band marks the semi-major axis of the orbits of the returning stellar debris, which is approximated by Kepler's third law assuming the energy distribution of the gas is frozen in at the time of disruption \citep[see equation 9 in][]{2014ApJ...783...23G}. The green shaded area shows the photospheric radius from the \texttt{MOSFiT} fit. Both the blue and green bands are bounded by a $1\sigma$ statistical error. The black crosses correspond to ASASSN observations in $V$-band reported on the \href{https://wis-tns.weizmann.ac.il/object/2018hyz}{Transient Name Server}, assuming that the underlying emission is a single temperature blackbody of $T_{\rm bb}=18,000$~K.
Right: Evolution of \Ha\ FWHM for a sample of TDEs. Except for ASASSN-14li, all other TDEs have remarkably similar FWHM of around $10^4$~\kms\ at early times. The error bars of AT~2018hyz denote the $1-\sigma$ uncertainty. Double-peaked signatures like the one seen for AT~2018hyz would be hard to identify if they are blended with the broad Gaussian emission lines in these TDEs. The dashed lines show the FWHM of the narrow \Ha\ component observed in AT~2018hyz ($\Delta t=$364 days) and AT~2018zr ($\Delta t=169$ days) at late times.
  \label{fig:fwhm}}
\end{figure*}

We measured the luminosity at each \swift\ epoch by integrating the best-fit blackbody spectrum of each UV-optical SED. We also calculated the emitting radius of the blackbody with the Stefan-Boltzmann law and plotted the evolution of the blackbody radius in the left panel of \autoref{fig:fwhm}. The size of the blackbody radius is tens of times larger than the pericenter distance of the disrupted star,
\begin{equation}
  R_{\rm p} \approx 7 \times 10^{12} \, \left ( \frac{M_{\rm bh}}{10^6 M_\odot} \right )^{1/3} \left ( \frac{R_\star}{R_\odot} \right ) \left ( \frac{M_\star}{M_\odot} \right )^{-1/3} {\rm ~cm},
\end{equation}

and shrinks with time.
The evolution of the blackbody temperature and radius is similar to that of previously studied TDEs \cite[e.g.]{2017ApJ...842...29H,2019ApJ...880..120H}.

Our follow-up photometric observations of AT~2018hyz were made after the peak of the UV/optical light curve. However, from the ASASSN data, the peak of the $g$-band light curve is estimated to occur around MJD$=58429$ \citep{2020arXiv200305469G}, which is very close to our first \swift\ observation on MJD$=58432$. We thus quote the luminosity measured from the first \swift\ epoch and estimate a radiated luminosity of $\gtrsim2.5 \times10^{44}$~erg~s$^{-1}$ at peak, which translates to an Eddington ratio of 0.6 for $M_{\rm bh}=3.5 \times 10^{6}$~\Msun. This places AT~2018hyz among one of a handful of TDEs radiating near the Eddington limit, such as ASASSN-14li and PS16dtm \citep{2017MNRAS.466.4904B, 2017ApJ...843..106B}, while the majority of previously studied TDEs (mostly X-ray faint) are found to radiate at $\sim$10\% level of the Eddington luminosity near peak \citep{2017ApJ...842...29H, 2017ApJ...843..106B}.

At such high luminosities, radiation pressure is expected to be large enough to increase the disk thickness and produce optically thick winds \citep{Shakura1973,2003MNRAS.345..657K,2005ApJ...628..368O}. Stream–stream collisions can also unbind a large fraction of shocked gas at the origin of disk formation, resulting in a collision-induced outflow \citep{2016ApJ...830..125J,2020MNRAS.492..686L}. These naturally give rise to an extended reprocessing envelope covering the X-ray emission site from a large solid angle \citep{1997ApJ...489..573L,2011MNRAS.410..359L,2018ApJ...859L..20D}. Similar to what is observed in other optically-discovered TDEs, the luminosity evolution of AT~2018hyz closely follows the classical mass fallback rate ($t^{-5/3}$) of the debris streams \citep{Holoien2014,2017ApJ...838..149A,2017ApJ...842...29H,2017ApJ...844...46B,Mockler2019, 2019ApJ...880..120H,2019ApJ...872..198V, 2020arXiv200305469G}. If these TDE flares are indeed powered by accretion onto the SMBH, this would require the disk to form promptly. Our analysis indicates that in AT~2018hyz, disk formation took place soon after the most-bound debris returned to pericenter, which we estimate to be about tens of days before the first detection (see \texttt{MOSFiT} results in \autoref{sec: bh mass}). However, as in all other events, the presence of an accretion disk has not been able to be discerned directly since the disk is likely hidden from our view due to orientation effects \citep{2018ApJ...859L..20D}.

\subsection{Spectroscopic Features}
The most notable feature in the spectra of AT~2018hyz is the double-peaked structure in the \Ha\ emission on day 51, which has two distinct peaks being displaced by 9320~\kms\ with the blue peak being stronger than the red one (\autoref{fig:disk_fit}). As will be detailed in the next section, the \Ha\ emission from day 51 to day 164 can be decomposed into disk and Gaussian components. On the other hand, the \Ha\ emission in the first spectrum, obtained on day 27, seems to lack an obvious double-peaked feature, and instead the entire line is characterized by a broad (${\rm FWHM} \approx 13,500$~\kms) flat-topped shape. This line profile is similar to that of the TDE AT~2018zr (PS18kh) and may arise from an elliptical disk with a low inclination angle \citep{2019ApJ...880..120H} or an expanding spherical outflow \citep{2019ApJ...879..119H}. We also note that the \ion{N}{3}+\ion{He}{2} emission, which is strongly detected in later epochs, has yet to emerge at $\Delta t = 27$~days. On the other hand, the low-ionization \ion{He}{1} $\lambda$5876 emission is strongest in the first spectrum.

A strong absorption feature at 5430~\AA\ is only observed in the day-51 spectrum but not in the other epochs. We identify it as  \ion{Na}{1}~D blueshifted by 23,600~\kms\ similar to blueshifted absorption features detached from their rest wavelengths that have been observed in other TDEs \citep[e.g.,][]{2019ApJ...879..119H}. We have also considered a highly blueshifted \ion{He}{1} $\lambda$5876 absorption to be responsible for this absorption feature. However, this possibility is ruled out since we do not detect blueshifted absorption from \ion{He}{1} $\lambda$3889, which is expected to be even stronger based on photoionization.

\subsection{Fitting \texorpdfstring{H$\alpha$}{ha} emission with the Elliptical Disk Model} \label{sec: e_disk_model_fit}

Following tidal disruption, the infalling material does not produce an TDE flare immediately. First material must circularize and form an accretion torus. There is no direct evidence supporting that the luminosity of a TDE follows the debris fallback rate that is not known a priori. This assumption, however, if true, would suggest that debris circularization must occur relatively quickly \citep{2015ApJ...809..166G,Mockler2019}. The bound orbits are very
eccentric and the orbital semi-major axis of the most tightly bound debris is
\begin{equation}
    a \approx 10^3 \left ( \frac{M_{\rm bh}}{10^6 M_\odot} \right )^{-1/3} \left ( \frac{R_\ast}{R_\odot} \right ) \left ( \frac{M_\ast}{M_\odot} \right )^{-2/3} R_{\rm g},
\end{equation} where $R_{\rm g}\approx 1.5 \times 10^{11} \, (M_{\rm bh}/10^6 M_\odot)$~cm is the gravitational radius. If the gaseous debris suffered no internal dissipation, it would form a highly elliptical disk with a large spread in apocentric distances between the most- and least-bound orbits. After pericenter passage, the infalling gas is on orbits that can interact with the infalling streams \citep{2009ApJ...697L..77R, 2015ApJ...804...85S,2016MNRAS.455.2253B,Gezari2017a,2014ApJ...783...23G,2015ApJ...809..166G, Dai13, Dai15,2013PASJ...65...86H,2016MNRAS.458.4250S,2020MNRAS.tmp.1380B},
giving rise to angular momentum redistributing shocks. The debris raining down would then be able to settle into a disk structure. Motivated by the predicted debris stream evolution, we use the elliptical disk model from \citet{1995ApJ...438..610E} to fit the \Ha\ emission in our spectra. We note that \citet{2017MNRAS.472L..99L} adapted this elliptical disk model for TDEs by further assuming that the disk size and orientation are determined by the dynamics of the intersecting debris streams. Here we choose to use the original \citet{1995ApJ...438..610E} prescription to remain agnostic to the mechanism(s) setting the extent of the elliptical disk.

We first fit the \Ha\ emission in the 51-day spectrum, which has the most prominent double-peaked feature, with an elliptical-disk model. The line profile produced by an elliptical geometrically thin, optically thick relativistic Keplerian disk has been formulated by \citet{1995ApJ...438..610E}. We constructed the model line profile by computing the line flux, $F_X$, in their Equation~7 at each velocity/frequency grid, $X$, with numerical integration after specifying seven parameters. The seven elliptical disk parameters are the power-law dependence of the emissivity profile $q$ (scales as $\xi^{-q}$), inner ($\xi_1$) and outer radii ($\xi_2$), inclination $i$, intrinsic broadening parameter $\sigma_{disk}$, eccentricity $e$, and azimuthal angle in the plane of the disk $\phi$. The inner and outer radii are defined as the pericenter distance of the inner and outer ellipses expressed in units of gravitational radius ($R_{\rm g}$). The inclination is defined such that $i=0$ when the disk is viewed face-on. The disk orientation angle ($\phi$) is measured with respect to the LOS of the observer, where $\phi=0$ when the apocenter aligns with the observer. While the 7 parameters determines the line shape, the overall normalization is set by the product of emissivity and black hole mass ($\epsilon_0 M^2$). However, as mentioned in \cite{Chen1989}, this does not allow us to estimate either parameter since the model expresses radius in dimensionless terms.

The dependence of the elliptical disk parameters on the computed line shape is non-trivial. Despite this, certain trends can be observed if we vary one parameter at a time as shown in Figure 3 in \cite{1995ApJ...438..610E} and Figure 17 in \cite{2003AJ....126.1720S}. For example, the emission line broadens as the separation between the two peaks becomes larger when we increase the inclination angle (from face-on to edge-on). As the outer disk ($\xi_2$) increases, the contribution from slow-moving material becomes stronger, and thus the line width narrows. In contrast, the smaller the inner disk radius ($\xi_1$) is, the broader the emission line is and the more pronounced the Doppler boosting is such that the blue peak becomes noticeably stronger relative to the red peak. Varying the orientation angle of the disk ($\phi$) also changes the relative strength of the two peaks, and so in some cases the red peak can be seen stronger than the blue peak. Increasing the local broadening ($\sigma_{disk}$) smooths the peaks of the line profile.

\begin{figure}
\begin{center}
\includegraphics[width=3.5in]{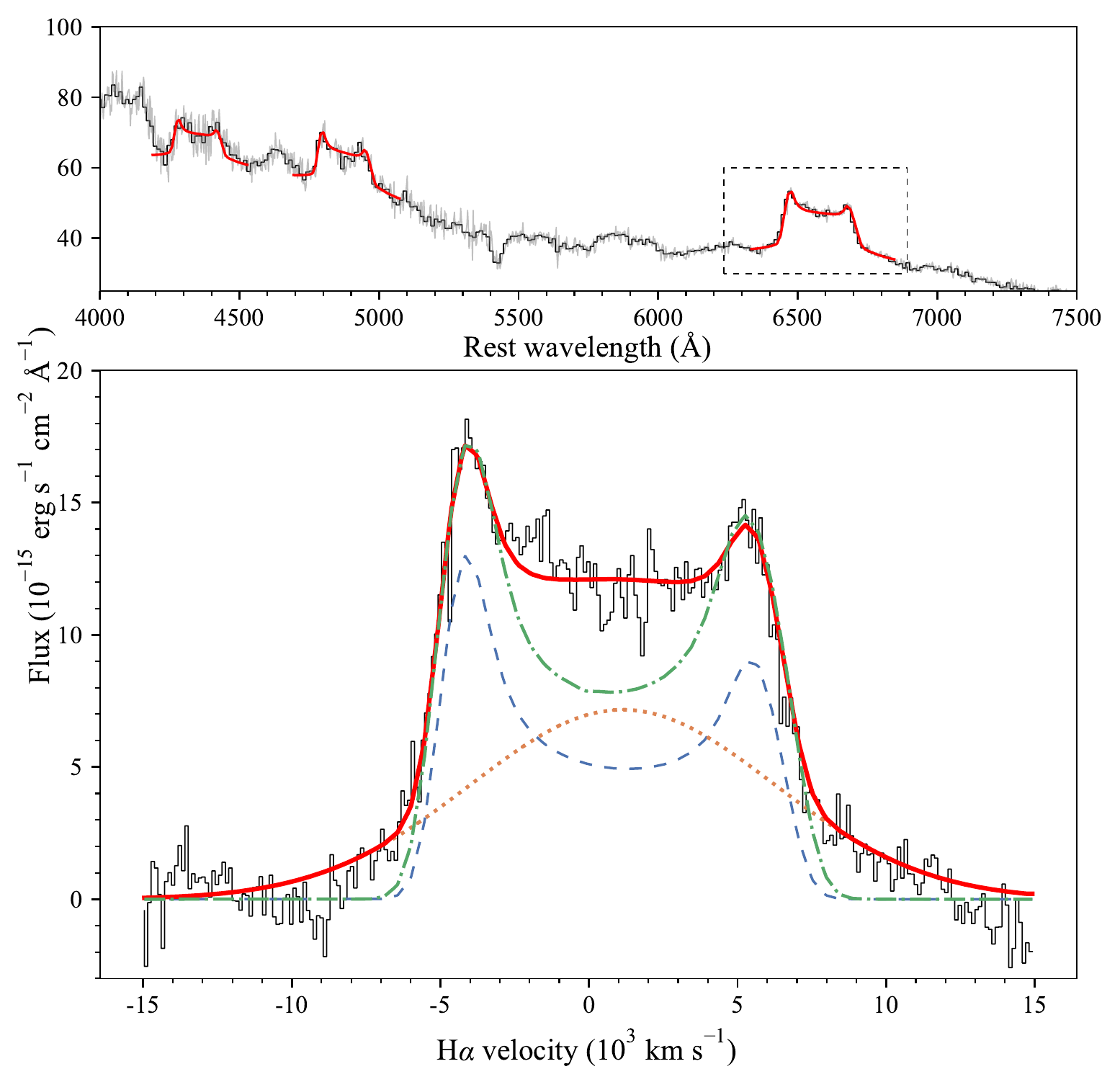}
\end{center}
\vskip -0.2in
\caption{$\Delta t=51$-day spectrum zoomed in on the \Ha\ emission line profile and best-fitting model.
The top and bottom panels are of the same flux scale.
The black histogram in the top panel shows the binned spectrum while the grey curve marks the original spectrum with host light subtracted. The red curves show the best-fitting two-component model derived from the \Ha\ profile (dashed box; enlarged in lower panel), rescaled by 1, 0.82, and 0.62 and shifted to the rest wavelength of \Ha, H$\beta$, and H$\gamma$ emission, respectively, on top of the local continuum. The local continuum is estimated by linearly interpolating the regions bracketing each emission line. The black curve in the lower panel shows the zoom-in view of the continuum-subtracted \Ha\ line profile. The solid red curve is the best-fitting two-component (disk + Gaussian) model with the parameters indicated in \autoref{tab:results}. The two components are shown separately in orange dotted line (Gaussian) and in blue dashed line (disk). The green line shows the best-fitting model with only the disk component by omitting the center part of the line flux.
It is clear that the disk-only model cannot reproduce the all flux near zero velocity.
  \label{fig:disk_fit}}
\end{figure}

We attempted fitting the observed \Ha\ emission line profile on day-51 with the elliptical disk model by minimizing $\chi^2$ in the following parameter space: $1 < q < 4$, $100 < \sigma_{disk} < 1,000$~km~s$^{-1}$, $0 \leq i < \frac{\pi}{2}$~rad, $0 < e < 1$, $0 \leq \phi < 2\pi$~rad, $100 < \xi_1 < 3,000$~$R_{\rm g}$, and $100 < \xi_2 < 10,000$~$R_{\rm g}$. We show the model parameters and the ranges of the flat priors in \autoref{tab:parameters}. We noticed that the best-fitting disk-only model does not capture the sharp peaks around the edges of the profile, but instead has broader peaks in order to match the high central flux.  Alternatively, if we attempt to fit only the edges of the profile by excluding the central data (with $|\Delta v| < 3000$~\kms), the best-fitting model, which is tabulated in \autoref{tab:results}, then lacks the flux seen at the center of the line profile (shown as the green line in \autoref{fig:disk_fit}). Clearly, a second component is needed to account for the central flux.

\begin{deluxetable*}{lcccc}
\centering
\tablecaption{Fitting parameters\label{tab:parameters}}
\tablehead{\colhead{Paramter} & \colhead{Notation} & \colhead{Unit} & \colhead{Min} & \colhead{Max}}
\startdata
\multicolumn{5}{c}{Elliptical Disk Component} \\
\hline
Emissivity power-law exponent  & $q$             &                   & 1              & 4               \\
Inner radius                   & $\xi_1$         & $R_g$           & $10^2$         & $3\times10^3$   \\
Outer radius                   & $\xi_2$         & $R_g$           & $10^2$         & $10^4$          \\
Inclination (rad)              & $i$             & rad               & 0              & $\frac{\pi}{2}$ \\
Intrinsic broadening parameter & $\sigma_{disk}$ & km s$^{-1}$     & $10^2$         & $10^3$          \\
Eccentricity                   & $e$             &                   & 0              & 1               \\
Orientation angle              & $\phi$          & rad               & 0              & $2\pi$          \\
\hline
\multicolumn{5}{c}{Gussian Component} \\
\hline
Amplitude                      & $A$             & Normalized flux & 0              & 1               \\
Velocity offset                & $\mu_0$         & km s$^{-1}$     & $-3\times10^3$ & $3\times10^3$   \\
Gaussian width                 & $\sigma_g$      & km s$^{-1}$     & 0              & $2\times10^4$
\enddata
\end{deluxetable*}

\begin{deluxetable*}{lcccccccccc}
\centering
\tablecaption{Elliptical-disk model constraints derived from the \Ha\ emission line profile\label{tab:results}}
\tablehead{
        & & \multicolumn{7}{c}{Disk} & \multicolumn{2}{c}{Gaussian} \\
        \hline
\colhead{Obs Date} & \colhead{Phase} & \colhead{$q$} & \colhead{$\sigma_{disk}$} & \colhead{$i$} & \colhead{$e$} &
\colhead{$\phi$} & \colhead{$\xi_1$} & \colhead{$\xi_2$}  & \colhead{$\mu_0$} & \colhead{FWHM}   \\
  &   \colhead{days} &   &  \colhead{\kms}    & \colhead{deg}    &     & \colhead{deg}    &  \colhead{$R_g$}  &  \colhead{$R_g$} & \colhead{\kms} & \colhead{\kms}}
\startdata
      2019-01-02 & 51 &  2.9$\pm$0.1 &    640 $\pm 70$   &   52$^{+13}_{-16}$  &   0.14$^{+0.06}_{-0.04}$      &   242$^{+27}_{-19}$   &   $1200^{+400}_{-500}$  & 2700 $^{+900}_{-1200}$  & - & - \\
      2019-01-02 & 51 & 2.0$\pm$0.5 &    570 $^{+90}_{-130}$   &   57$\pm13$       &   0.13$^{+0.05}_{-0.02}$   &   257$^{+27}_{-28}$   &   1500$^{+400}_{-500}$  & 2600 $^{+600}_{-900}$ & 990$^{+520}_{-450}$ & $11900^{+1050}_{-800}$ \\
      2019-03-15 & 120 &  2.0  &  450$^{+250}_{-170}$  &   68$^{+11}_{-18}$  &   0.10 $^{+0.1}_{-0.06}$ &   229$^{+91}_{-78}$ &   $1900^{+800}_{-700}$     & 3100$^{+900}_{-1000}$ & 660$^{+360}_{-300}$ & $10380^{+1070}_{-1540}$ \\
      2019-05-01 & 165 &  2.0 &   370$^{+180}_{-110}$   &   65 $^{+13}_{-18}$     &   0.09 $^{+0.1}_{-0.04}$ &   222$^{+77}_{-25}$ &   1800$^{+500}_{-600}$ & 2800$^{+700}_{-1000}$   & 140$^{+90}_{-80}$ & 8900$^{+300}_{-500}$  \\
\enddata
\tablecomments{The \Ha\ emission line on 02 Jan 2019 was fit with a disk-only model (first line) and a disk + Gaussian model (second line).  For all other epochs, results from only the disk + Gaussian model are shown. The uncertainties listed are the 16th- and 84th-percentile of the marginal posterior distribution for each parameter.}
\end{deluxetable*}

\begin{figure*}
\begin{center}
\includegraphics[width=6.5in]{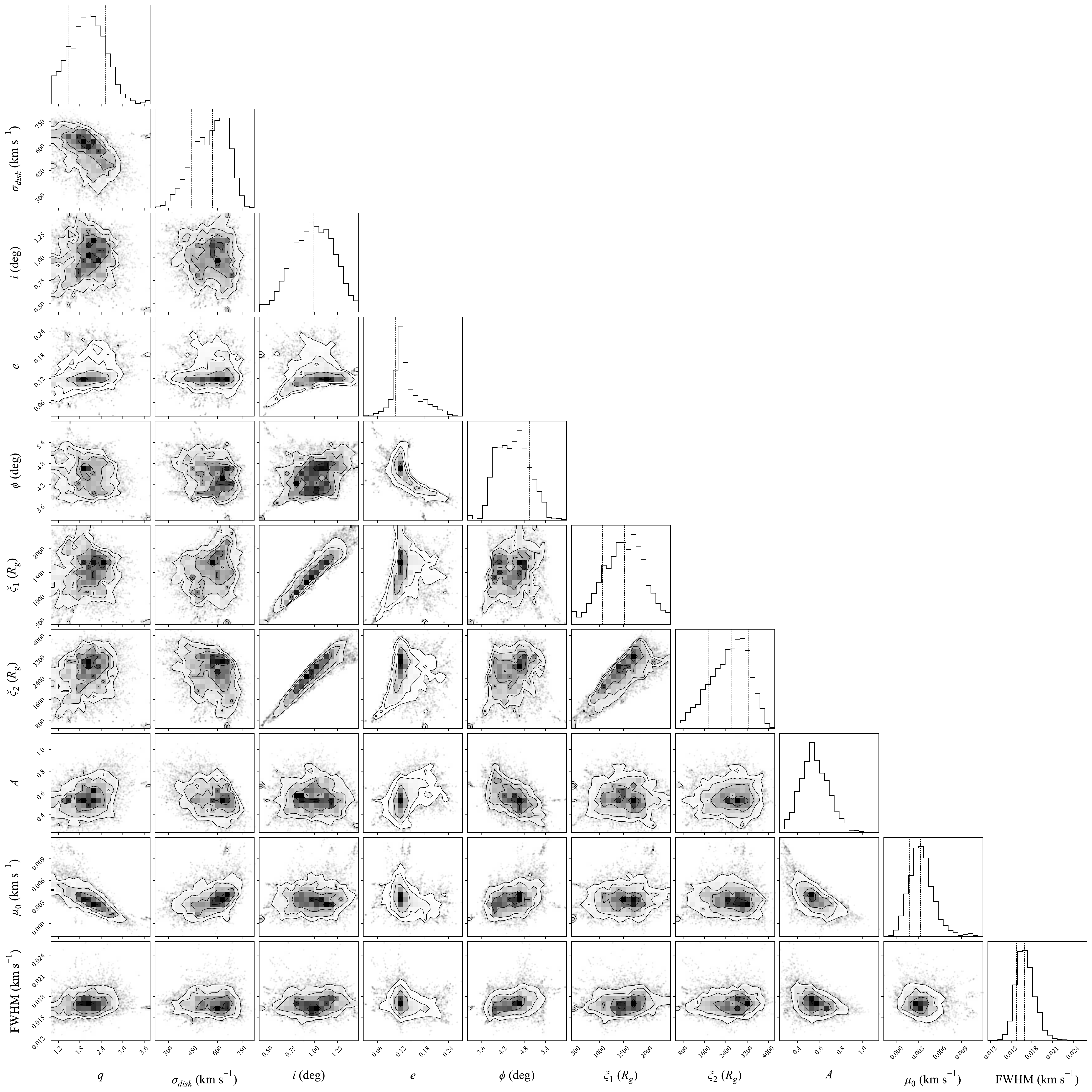}
\end{center}
\vskip -0.2in
\caption{Posterior distributions of parameters for the two-component elliptical disk + Gaussian model. The histograms at the top of each column show the marginalized posterior distribution for each parameter. The vertical dashed lines mark the quoted values in \autoref{tab:results}, which correspond to the 16th, 50th, and 84th percentile, respectively. For each parameter pair, the joint posterior distribution is shown as a 2D histogram where the contours correspond to the 16th, 50th, and 84th percentile.}
  \label{fig:posterior_dist}
\end{figure*}

In order to match the full line profile, we add a Gaussian component to the disk model. We smoothed the data of the \Ha\ emission with a boxcar function with a length of $\sim$5~\AA\ to erase fine structures that cannot be captured by the emission-line model. Using the \texttt{emcee} package \citep{emcee}, we sampled the posterior distribution of a total of ten parameters (the seven disk parameters listed above and an additional three parameters that describe the Gaussian: amplitude, center ($\mu_0$; velocity offset from the systemic velocity of the host galaxy), and width ($\sigma_g$; standard deviation) using flat priors over the aforementioned ranges for the disk parameters and over the following ranges for the Gaussian component: $-3,000 < \mu_0 < 3,000$~km~s$^{-1}$, and $0 < \sigma_{g} < 20,000$~km~s$^{-1}$ (\autoref{tab:parameters}). The model parameter $\sigma_g$ can be translated to the FWHM of the Gaussian profile via the relation FWHM$=2\sqrt{2\textrm{log}2}\sigma_g$. We normalized our flux to range between 0 and 1 and use this same range as the flat prior for the amplitude of the Gaussian. The posterior distributions of the model parameters are shown in \autoref{fig:posterior_dist}. The MCMC approach provides a more general description of the model parameters, instead of forcing them to be Gaussian distributed as in traditional fitting methods. In addition, we can also visualize the correlation between different parameters in the joint distributions, such as that between $i$, $\xi_1$, and $\xi_2$, which intricately control the overall width of the disk profile. From this, we derived the parameters for the two-component (disk-plus-Gaussian) model and present the derived parameters in \autoref{tab:results} and compare the best-fitting model to the spectrum in \autoref{fig:disk_fit}. The blue \Ha\ emission peak being relatively stronger than the red peak can be explained by Doppler boosting in the scheme of an axisymmetric accretion disk, though we note that the relative strength of the two peaks is orientation-dependent in elliptical disks.

To summarize, the day-51 spectrum is best described by two distinct spectral components: a non-disk broad Gaussian component with a centroid velocity close to the systemic velocity of the host (similar to that observed in all other TDEs), and a prominent disk/double-peak component (see \autoref{fig:disk_fit}).
The \Ha\ line profile in the day-120 and day-164 spectra still shows a double-horned shape, despite being less prominent relative to the Gaussian component than that in the $\Delta t = 51$ days spectrum. However, because the disk parameters are the most constrained by the shape and strength of the double-peaked structure, the dominance of the Gaussian component at later stages means that performing a fully independent fit for these epochs would result in loosely constrained model parameters. Therefore, when fitting the \Ha\ emission with the MCMC in the later two epochs, we fixed the surface emissivity to the best-fit value ($q=2$) from day-51. This value is between the emissivity index of local dissipation of gravitational energy \citep[$q=1.5$, such as seen in cataclysmic variable disks;][]{1991ApJ...374L..55H}, and that of a disk illuminated by an isotropic ionizing source \citep[$q=3$;][]{1989ApJ...339..742C,2012MNRAS.424.1284W}.
The rest of the parameters were set to have the same flat priors as above. Although the disk parameters are less constrained in the two later epochs (\autoref{tab:results}), the inclination angle, and inner/outer disk radii are still consistent with the fit for the day-51 spectrum. We find a decreased relative disk contribution to the flux over time. On day-51, the integrated flux of the disk is comparable to the Gaussian component but only 28\% of the Guassian component in the day-120 spectrum. We plot the size of the elliptical disk and the broad-line region (BLR; derived from the Gaussian component) in all three epochs along with the photospheric radius derived from \swift\ photometry in the left panel of \autoref{fig:fwhm}. In the right panel of \autoref{fig:fwhm}, we show the evolution of the best-fit FWHM of the Gaussian component compared to the evolution of \Ha\ width in other TDEs. The width and the evolution of this Gaussian component are similar to that seen in other optical TDEs \citep{2017ApJ...842...29H}.

We did not attempt fitting the first spectrum (day-27) since the flat-topped line shape without visible blue- and red-shifted peaks leaves our model parameters poorly constrained. However, it is likely that the \Ha\ emission from the $\Delta t=27$~d spectrum also consists of the same disk and the Gaussian component given their FWHMs do not vary significantly. The double-peaked disk feature on $\Delta t=27$~d may be buried due to a relatively weaker disk contribution reletively to the Gaussian component.
We did not fit the \Ha\ in the $\Delta t=117$~d spectrum because of its proximity in time to the $\Delta t=120$~d spectrum. As can be seen in \autoref{fig:opt_spectra}, the line shape barely changed in this three-day period. We also omitted fitting the $\Delta t=199$~d spectrum because the spectral coverage does not extend redward enough to cover the entire \Ha\ emission. The \Ha\ in the late-time spectrum became a narrow emission line thus we did not attempt fitting it with an elliptical accretion disk.

From our analysis, we find that a quasi-circular disk ($e\approx0.1$) with a mean semi-major axis ($a\equiv\frac{\xi_1+\xi_2}{2(1-e)}$) of 2400$R_g$ ($\approx1.2\times10^{15}$~cm) is responsible for the double-peaked \Ha\ emission in AT~2018hyz. Assuming Keplerian motion, the velocity separation of the two \Ha\ peaks ($V_{obs}$) at $\Delta t=51$ days corresponds to a radius of $R/R_g=(\frac{2c\sin{i}}{V_{obs}})^2=2900$, which is consistent with the results from our model fitting.

Although the central part of the \Ha\ emission overlaps with the telluric B-band, we were able to recover a smooth emission profile after correcting for the telluric absorption (\autoref{fig:telluric}). As a sanity check, we note that the other Balmer lines (H$_\beta$ and H$_\gamma$) observed at the same time (day-51) also have the same double-peaked line profile. Both \Hb\ and \Hg\ can be successfully described by the same disk + Gaussian model (see \autoref{fig:disk_fit}) rescaled and laid on top of a local continuum, lending further support for the realness of the double-peaked emission. As shown by \cite{2020arXiv200305470S}, the velocity profiles of \Ha\ and \Hb\ of AT~2018hyz are broadly consistent throughout the first 5 months. Therefore, the scaling factor derived for the \Hb\ emission is directly measuring the Balmer decrement (\Ha/\Hb $\approx1.2$). The flat Balmer decrement (relative to the typical case B ratio of 3 for AGNs) suggests that the production of the Balmer emission is dominated by collisional excitation rather than photoionization at the time of the observation. In an independent work on AT~2018hyz, \cite{2020arXiv200305470S} also derived a \Ha/\Hb\ of 1.5 and pointed out that such a ratio can be achieved in the outer optically thin part of a Cataclysmic Variable (CV) disk (Williams 1980). However, the same mechanism that produces the flat Balmer decrement cannot explain the appeareance of \ion{He}{1} or \ion{He}{2}.

\begin{figure*}
\begin{center}
\includegraphics[width=3.2in]{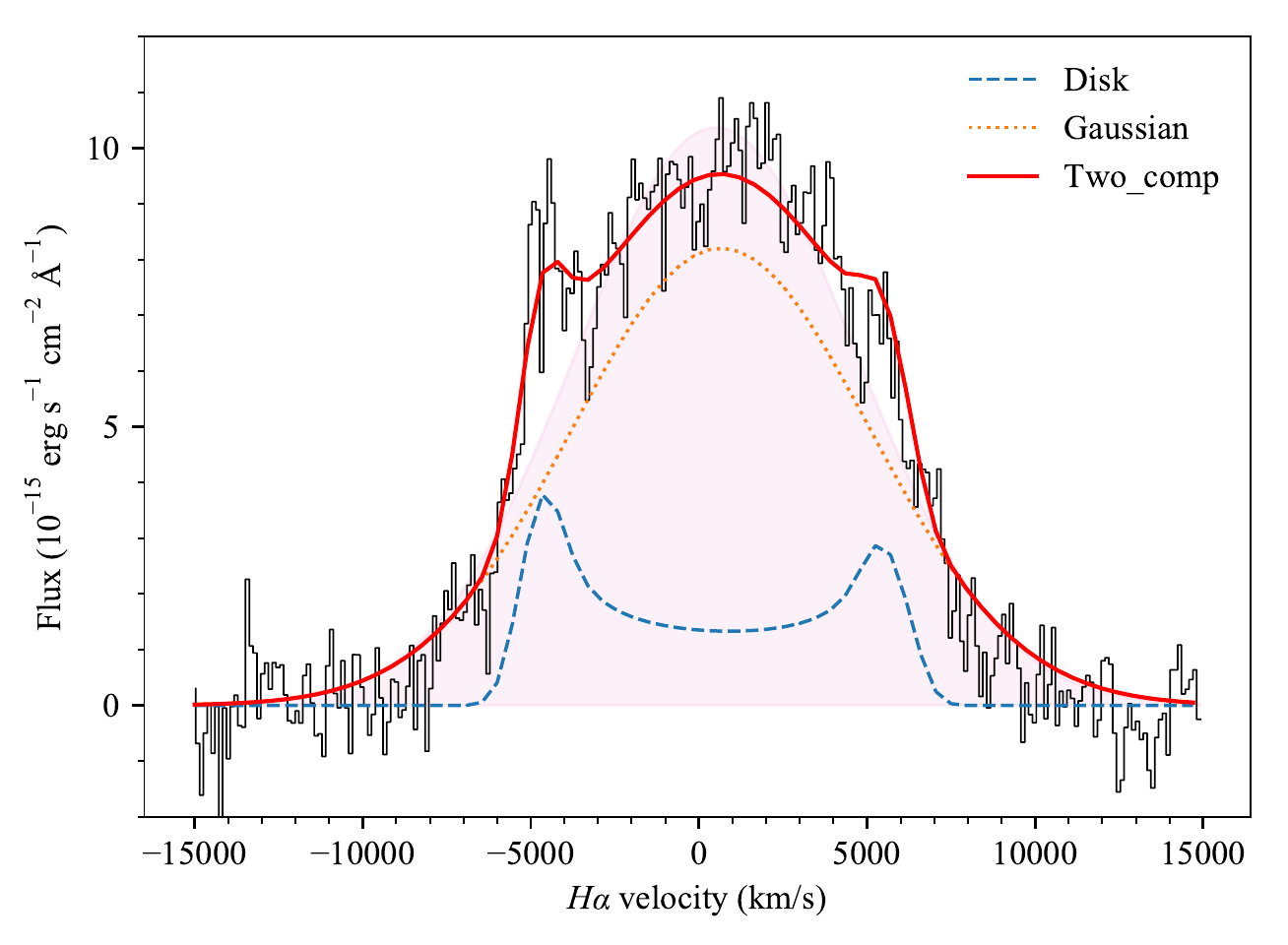}
\includegraphics[width=3.2in]{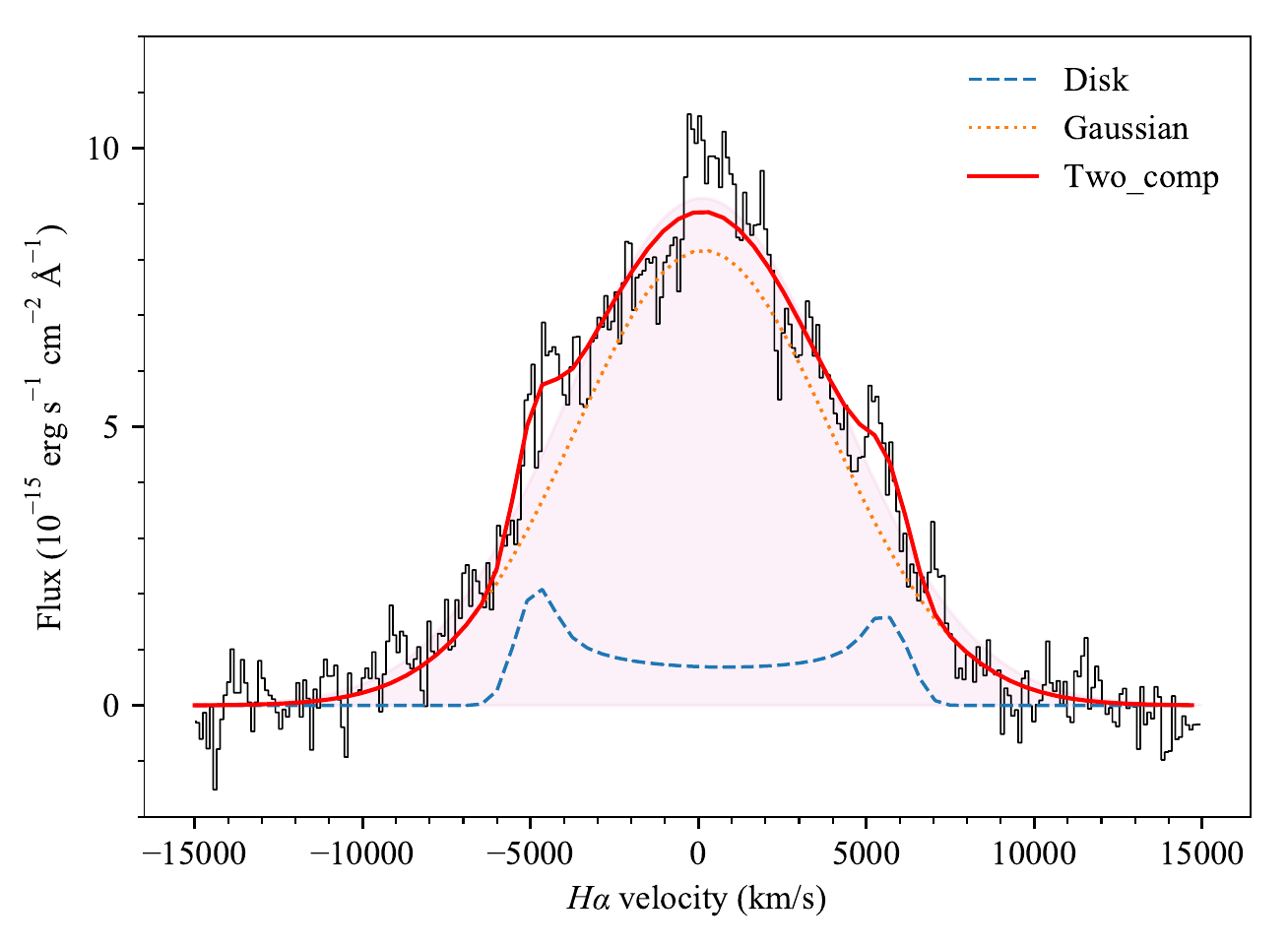}
\end{center}
\vskip -0.2in
\caption{Best-fitting two-component models (red curves) for the \Ha\ emission of AT~2018hyz (black curves) from the $\Delta t=120$~days (Left) and $\Delta t=165$~days (Right) spectra. The disk and Gaussian components are shown as blue and orange curves, respectively. The pink shaded area indicates the best-fitting single-Gaussian model to the entire emission line profile, indicating that this model is insufficient at matching the peaks near $|v| = 5000$~\kms. Despite having notable disk emission, the line profile at these late phases are relatively similar to the flat-topped or single-Gaussian line profiles typically seen in TDEs.
  \label{fig:late_time_disk_fit}}
\end{figure*}

\section{Discussions}
\label{sec:discussion}

\subsection{Alternative origin for the double-peaked line profile}
Accretion disk emission is not the only model for double-peaked emitters. In a TDE environment, another plausible scenario that may result in a double-peaked line profile is the emission from a bipolar outflow. If the redshifted wind component receding from us is not obstructed from our line of sight, one can possibly observe double-peaked emission line profiles. This scenario cannot be completely ruled out even though radio emission from AT 2018hyz was not detected \citep{2018ATel12218....1H}. The strength of the radio signal also depends on the density of the circumnuclear gas and therefore TDEs with winds do not necessarily produce strong radio emission \citep{Alexander2016}. From the current TDE sample, there is no association between radio-detected events and the appearance of double-peaked emission lines.

Nevertheless, it is still unclear whether TDEs can produce outflows with the required geometry, density, and velocity to explain the \Ha\ line profile seen in AT~2018hyz, especially if prompt disk formation has not happened in this event. Simulations have shown that stream-stream collisions in TDEs tend to produce winds that are more or less spherical rather than bipolar \citep[e.g.][]{2016ApJ...830..125J}. Assuming an accretion origin for the TDE light, \cite{2018ApJ...855...54R} showed that a spherical wind can give rise to an asymmetrical flat-topped line profile, similar to that observed in AT~2018zr \citep{2019ApJ...879..119H}. Winds driven by the radiation pressure in super-Eddington disks can produce winds that look bipolar when viewed close to the polar axis. However, such winds are optically thick, therefore are more likely to result in P-Cygni features or an asymmetrically broadened/flat-topped emission line.

Given all the evidence, we find the bipolar outflow model less appealing. On the other hand, the disk-plus-Gaussian model provides an excellent fit to the data with model parameters that are physically reasonable. Given current observational evidence, a Keplerian accretion disk is more likely to be responsible for producing the observed double-peaked \Ha\ emission in AT~2018hyz.

\subsection{Angular Momentum}

Assuming the disrupted star was initially on a parabolic orbit, the total angular momentum of the star at pericenter ($R_{\rm p}$) is $\ell_\star = M_\star\sqrt{2GM_{\rm bh}R_{\rm p}}$. For the stellar debris to settle into a circular disk, the bulk of the mass must fall in a disk with a radius of $\sim$2$R_{\rm p}$ according to the conservation of specific angular momentum.
At first glance, the fact that the size of the elliptical disk ($r\approx1.2\times10^{15}$~cm) is orders of magnitude larger than $R_{\rm p}$ poses a physical problem. However, as pointed out in \citet{2019ApJ...880..120H}, this may be resolved if the mass of the line-emitting gas is small compared to the mass of the initial star. As such, there exist several ways to transport a small amount of mass (e.g., through shocks, advection through wind, etc.) and redistribute the angular momentum quickly. We find it unlikely for the disk to spread to such a large distance through viscous processes in the observed timescale, given that the viscous timescale at the mean semi-major axis is much longer.
\begin{equation} \label{eq1}
\begin{split}
t_{\rm visc} & = \alpha^{-1} (H/R)^{-2} t_{\rm dyn}\\
 & = 158 \left(\frac{\alpha}{0.1}\right)^{-1} (H/R)^{-2}~{\rm days}
\end{split}
\end{equation}
Depending on the thickness of the disk at this distance, the viscous timescale is on the order of one to a thousand years.

We can estimate the mass of the line-emitting part of the disk by considering the conservation of total angular momentum. The total angular momentum of the line-emitting region of the disk is
\begin{equation}
\begin{split}
    \ell_{disk} &= M_{disk} \sqrt{G M_{bh} a(1-e^2)} \\
    &= M_{disk} \sqrt{G M_{bh} \xi (1+e)},
\end{split}
\end{equation}

where $a$ is the semi-major axis, $\xi$ is the pericenter distance, and $e$ is the eccentricity. In this calculation, we evaluate the angular momentum at $\xi_2$. Assuming the bound stellar debris has a mass of $\sim${}$0.5 M_\star$, we require $\ell_{\rm disk} \lesssim 0.5\ell_\star$. From the $\Delta t=51$~d observation, we find an upper limit

\begin{equation}
    M_{\rm disk} \lesssim 0.5M_\star \sqrt{\frac{R_{\rm p}}{\xi_2(1+e)}} = 0.04 M_\star.
\end{equation}

Our analysis of the \Ha\ emission implies that the mass of the extended disk is small compared to the mass of the disrupted star and that significant transport of angular momentum has taken place in order for the disk to spread to those large distances.


Our observations show that the accretion disk in AT~2018hyz extends out to $\sim${}$100R_{\rm p}$. The smooth, uniform decline in the UV and optical bands suggests no change in the dominant emission mechanism during the monitoring period and hints at the importance of accretion operating effectively in the entire duration of the TDE.

Recent numerical simulations that focus on the disk formation process have found that shocks induced by self-intersection of the debris streams are capable of redistributing angular momentum \citep{2015ApJ...804...85S,2016MNRAS.455.2253B,2016MNRAS.458.4250S,2020MNRAS.tmp.1380B}. \cite{2020MNRAS.tmp.1380B} found a relatively circular ($e\approx$0.2), geometrically thick ($H/R\approx1$) disk extending to $\sim 300R_g$ may be achieved efficiently in 0.5$t_{mb}$ (fallback time of the most-bound debirs) following numerous secondary shocks generated near the self-crossing point. Within the thick disk, the gas angular momentum is sub-Keplerian, which implies that the measured linewidth at the same distance would be narrower than that assuming Keplerian rotation. The fact that our derived disk sizing being a few factors larger does not necessarily contradict the results from these theoretical works as we may be looking at different parts of the disk. Previous works have shown that the \Ha\ photosphere lies near the electron scattering surface \citep{Roth2016}, which has a size of $\sim1000 R_g$ at sight lines away from the funnel \citep{2018ApJ...859L..20D}. The compact, puffed up TDE disk as described by \cite{2018ApJ...859L..20D} and \cite{2020MNRAS.tmp.1380B} would be opaque to the \Ha\ emission. Our observations provide evidence for the presence of a thin, nearly circular \Ha\ emitting disk, which may or may not be attached to a thicker disk in the center.

\subsection{Non-disk Gaussian Component}

Theories and observations have indicated that a radiatively-driven wind may be ubiquitous among TDEs during their early phases where the debris fallback rate is thought to be ``super-Eddington'' \citep{Strubbe2009,2011MNRAS.415..168S,2018ApJ...859L..20D, Alexander2016,2019ApJ...879..119H}. Such a wind naturally provides a physical origin for the BLR that is likely to be responsible for the broad Gaussian component observed in AT~2018hyz and in the other optical TDEs. According to our fit, the disk component contributes half of the total \Ha\ line flux. The Gaussian component has a FWHM of $12,000$~\kms\ which corresponds to a virial radius of $\sim${}$8.9\times10^{14}$~cm, or $\sim${}$110 R_{\rm p}$ (=1700$R_g$), and slowly narrows ($\leq$30\%) over the duration of spectroscopic monitoring. Both the line width and its lack of evolution in AT~2018hyz are similar to the broad-line emission properties of other TDEs (\autoref{fig:fwhm}).

Broad lines are also present in many AGNs and they are thought to be produced by gas above and below the disk plane at distances from light-hours to light-years away from the SMBH. In a TDE, the matter distribution that ensheathes the SMBH is likely produced quickly by a wind that is supported by a combination of radiation pressure and rotation. Accretion then proceeds through the mid-plane. In a geometrically thin, optically thick disk, the \Ha\ line emitting portion of the disk corresponds to an effective temperature of about 8,000~K. Changes in the properties of the disk, such as in the rate of mass supply, naturally result in changes in the line emission on an orbital timescale on the order of weeks. This is in agreement with the timescale at which the double-peak line profiles are observed to change.

\subsection{Bowen emission lines}
The appearance of the feature around 4650\AA\ on day-51 is either due to a blueshifted \HeII\ or a \ion{N}{3}$\lambda4641$+\HeII\ emission complex. This feature persisted in all of the later spectra and provides additional evidence for ongoing accretion \citep{2019ApJ...887..218L}. In particular, the \ion{N}{3} complex is known to be produced by the Bowen fluorescence mechanism \citep{1935ApJ....81....1B}. The mechanism can be understood as follows. First, the ionization of \ion{He}{2} at an energy of 54.4eV produces \HeII\ as well as the \ion{He}{2} Ly$\alpha$ photons upon recombination. The \ion{He}{2} Ly$\alpha$ line at $\lambda=303.78$\AA\ has a energy that is remarkably similar to the \ion{O}{3} transitions ${\rm 2p^2\,{}^3P_2-2p3d\,{}^3P^{\circ}_2}$ and ${\rm 2p^2\,{}^3P_2-2p3d\,{}^3P^{\circ}_1}$. The radiative excitation of the \ion{O}{3} resonance line and the subsequent cascades produce several \ion{O}{3} lines, one of which then in turn excites the \ion{N}{3} doublet $\lambda\lambda374.434, 374.441$. The end result of this is the production of many \ion{O}{3} and \ion{N}{3} emission in the optical wavelength range of $3000-4650$ \AA. Within the wavelength coverage of our spectra, we also looked for Bowen emission \ion{O}{3}$\lambda$3760 and \ion{N}{3}$\lambda\lambda4097,4104$ in our spectra. Because the \ion{N}{3}$\lambda\lambda4097,4104$ lines overlaps with H$\gamma$ we cannot confidently establish the detection on day-51. We note that this feature became buried in the continuum in later spectra. The \ion{O}{3}$\lambda$3760 line is marginally detected in the $\Delta t=164$ days spectrum, when the \HeII\ complex is the strongest.

In TDEs with a low X-ray-to-optical ratio like AT~2018hyz, the observed blackbody spectrum generally could not produce enough ionizing photons to trigger the Bowen mechanism. For example, even when assuming a generous 100\% energy conversion efficiency, the blackbody emission in AT~2018hyz only produces $\lesssim${}$10^{32}$~erg~s$^{-1}$ of luminosity below the ionization edge of \ion{He}{2} (228~\AA) and thus cannot account for the observed \ion{N}{3}+\ion{He}{2} luminosity of $3\times10^{39}$~erg~s$^{-1}$. Therefore, an additional extreme UV or X-ray source, likely an accretion disk, is required.

\cite{2020arXiv200101409V} classified the spectra of TDEs into 3 different spectral classes, namely TDE-H , TDE-Bowen, and TDE-He that are found to be associated with distinctive rise time distributions. These three spectral classes are defined based on the species of the broad emission lines present in the spectra. The spectrum of TDE-Bowen's shows both broad \HeII\ and Balmer emission. TDE-H's and TDE-He's have spectra that feature only broad hydrogen and \ion{He}{2} emission, respectively. Under this classification scheme, AT~2018hyz is shown to transition from TDE-H to TDE-Bowen in less than 2 months. The origin of this transition in spectral class in AT~2018hyz is still largely unknown yet has been reported in another TDE, ASASSN-14ae \citep{Holoien2014,2016MNRAS.462.3993B}. \cite{2016MNRAS.462.3993B} attributed this change to the hardening of the radiation field at late time. The composition of the fallback stellar material may also affect changes in the helium-to-hydrogen ratio in AT~2018hyz. \cite{2019ApJ...882L..25L} showed that the helium abundance can start to rise months after the disruption. However, we note that a detailed radiative transfer calculation, which is beyond the scope of this paper, is required to demonstrate how the change in abundance translates into the observed line ratio.

\subsection{Late-time spectroscopic evolution}
In the last spectroscopic epoch obtained on $\Delta t=364$~d, the spectrum almost fades back to the host level. As can be seen from \autoref{fig:opt_spectra}, the only features in the residual spectrum is a narrow \Ha\ emission ($920\pm60$~\kms) at a systemic velocity of the host galaxy and a weak \ion{N}{3}+\ion{He}{2} complex (red dashed line in \autoref{fig:fwhm}). As shown in \autoref{fig:fwhm}, it is common to see emission lines in TDE spectra narrow with time \citep[e.g.][]{2016MNRAS.462.3993B,2017ApJ...842...29H} yet not with such a significant reduction (factor of ten) in width. Comparing to the Gaussian component of the spectra observed at earlier epochs, this change in line width suggests that the production site of \Ha\ emission has moved from $\sim${}$10^{15}$~cm to $\sim${}$10^{17}$~cm.

A similar evolution of \Ha\ emission at late time is seen in AT~2018zr \citep[see the pink dashed line in \autoref{fig:fwhm};][]{2019ApJ...879..119H}. This narrow component is likely to arise from the circumnuclear gas that lies much farther out being photoionized by the TDE flare. However, since very few TDEs have spectra a year after peak brightness, we cannot yet verify whether this is trend is typical.

\subsection{Double-peaked line profiles in TDEs}

\citet{2017MNRAS.472L..99L} described a peculiar \Ha\ emission feature in the TDE PTF09djl with two peaks being separated by $3.5\times10^4$~\kms, suggesting it was caused by an elliptical disk with extremely high inclination and eccentricity. However,
this feature was only seen in a set of three low-S/N spectra where the \Ha\ feature falls in several telluric bands and therefore accurate continuum determination cannot be achieved. Because of these complications, PTF09djl does not show a clear and robust detection of a disk.

The derived disk properties of AT~2018hyz are also quite different from that of PTF09djl. AT~2018hyz has a fairly circular disk with a moderate-to-high inclination angle while the purported PTF09djl disk is characterized by extremely high ellipticity ($e \approx 0.97$) and a high inclination angle ($i\approx88^{\circ}$). In addition, a significant portion of the \Ha-emitting disk in PTF09djl (extending from $30 R_{\rm g}$ to $400 R_{\rm g}$) while the disk of AT~2018hyz locates at 1000--$3000 R_{\rm g}$. Our results are derived from the original \cite{1995ApJ...438..610E} prescription, which approximated photon geodesics under the weak-field limit, thus the validity of the model breaks down at $\xi<100R_g$. We note that \cite{2017MNRAS.472L..99L} applied analytical forms of photon geodesics under strong gravity regime, which is methodologically different from our calculation. The best-fit disk parameters presented by \cite{2017MNRAS.472L..99L} suggest that the \Ha\ emission in PTF09djl is produced by the debris streams on highly eccentric orbits, which remained the same size for at least a duration of 2 months. Their analysis indicates that the stellar debris are unable to circularize by dissipating orbital energy.

AT~2018hyz provides a more compelling case for a disk, consistent with the classical picture where a disk forms efficiently following a TDE, though we note that recent simulations and analytical calculations have found disk formation to be inefficient in several regions of the parameter space \citep[e.g.,][]{Dai15,2015ApJ...804...85S,2016MNRAS.461.3760H,2016MNRAS.455.2253B,2017MNRAS.tmp..118S}. However, even if circularization is efficient in most TDEs, as deduced for example by the detection of Bowen emission lines in their early phases \citep{2019ApJ...873...92B,2019ApJ...887..218L}, double-peaked features have not been unambiguously observed in other events. It is likely that this signature is missed for configurations in which the accretion disk is less inclined, which results in them being much weaker relative to the more isotropic, broad Gaussian component (see the \Ha\ line profiles in later epochs \autoref{fig:late_time_disk_fit}). The disk of AT~2018hyz is deduced to be highly inclined with $i=57^{\circ} \pm 13$. A more face-on disk (with a lower inclination angle) may result in a more flat-topped line profile as the separation between the two peaks diminishes \citep[e.g., the \Ha\ emission in AT~2018zr could arise from a low-inclination disk;][]{2019ApJ...880..120H}. Despite the disk contribution being weaker at later epochs for AT~2018hyz, the disk inclination angle remains high and shows no significant precession. On the other hand, the eccentricity declines slightly, hinting that energy dissipation possibly by shocks continues to operate efficiently.

\section{Conclusions}
\label{sec:conclusion}

To summarize, we find that the observations of AT~2018hyz are consistent with the classical theoretical prediction in which TDE flares are powered by a newly-formed accretion disk, which in turn is responsible for producing the broad-line region. In the context of TDE debris stream evolution, we find the first strong observational evidence for prompt circularization following the disruption of a star. This accretion is accompanied by vast amounts of energy release that results in a wide range of key elements of the flow patterns, all necessary to explain the salient properties of TDEs and are highly-reminiscent of the well-studied phenomenology of steadily-accreting AGN. These are an accretion disk, responsible for generating the bulk of the power, and an extended gaseous envelope, responsible for reprocessing the radiation as well as producing the broad line features. Whereas AGN are supplied by a steady stream of fuel for thousands of years, TDEs like AT~2018hyz offer a unique opportunity to study a single SMBH under a set of conditions that vary dramatically over months. For this reason, studying objects like AT~2018hyz provide the firmest hope of understanding the physics of accretion disks around SMBHs for a wide range of accretion conditions.

\acknowledgments

T.H.\ thanks Prof.\ Michael Eracleous for discussion on numerical concerns when implementing the elliptical disk model, Dr.\ Qian Wang for numerical and algorithmic advice to greatly speed up computation, and Dr.\ Martin Gaskell for discussion on the absorption and emission features in the spectra. The authors would like to thank anonymous referee for suggestions that greatly
improved the clarity of the paper. The UCSC transient team is supported in part by NSF grant AST-1518052, NASA/{\it Swift} grant 80NSSC19K1386, the Gordon \& Betty Moore Foundation, the Heising-Simons Foundation, and by a fellowship from the David and Lucile Packard Foundation to R.J.F.  K.A.A., J.L.S., E.R.-R., and B.M.\ are supported by the Danish National Research Foundation (DNRF132), the Heising-Simons Foundation and NSF grant AST-161588. J.L.D. is  supported  by  the  GRF  grant  from the Hong Kong government under HKU 27305119. M.R.S. is supported by the National Science Foundation Graduate Research Fellowship Program Under Grant No. 1842400.

Parts of this research were supported by the Australian Research Council Centre of Excellence for All Sky Astrophysics in 3 Dimensions (ASTRO 3D), through project number CE170100013.

Research at Lick Observatory is partially supported by a generous gift from Google.

Some of the data presented herein were obtained at the W.\ M.\ Keck Observatory, which is operated as a scientific partnership among the California Institute of Technology, the University of California and the National Aeronautics and Space Administration. The Observatory was made possible by the generous financial support of the W.\ M.\ Keck Foundation.  The authors wish to recognize and acknowledge the very significant cultural role and reverence that the summit of Maunakea has always had within the indigenous Hawaiian community.  We are most fortunate to have the opportunity to conduct observations from this mountain.

Based on observations obtained at the Southern Astrophysical Research (SOAR) telescope, which is a joint project of the Minist\'{e}rio da Ci\^{e}ncia, Tecnologia, Inova\c{c}\~{o}es e Comunica\c{c}\~{o}es (MCTIC) do Brasil, the U.S. National Optical Astronomy Observatory (NOAO), the University of North Carolina at Chapel Hill (UNC), and Michigan State University (MSU).

This work includes data obtained with the Swope Telescope at Las Campanas Observatory, Chile, as part of the Swope Time Domain Key Project (PI: Piro, Co-Is: Drout, Phillips, Holoien, French, Cowperthwaite, Burns, Madore, Foley, Kilpatrick, Rojas-Bravo, Dimitriadis, Hsiao). We wish to thank Swope Telescope observers Jorge Anais Vilchez, Abdo Campillay, Nahir Munoz Elgueta and Natalie Ulloa for collecting data presented in this paper.

\software{photpipe imaging and photometry pipeline \citep{Rest2005,Rest2014}, hotpants \citep{HOTPANTS}, DoPhot \citep{Schechter93}, XSPEC (v12.10.1f; Arnaud 1996), PyRAF (Science Software Branch at STScI 2012), MOSFit \citep{2018ApJS..236....6G,Mockler2019}, emcee package \citep{emcee}}


\bibliography{tde}{}
\bibliographystyle{aasjournal}

\appendix

\setcounter{table}{0}
\renewcommand{\thetable}{A\arabic{table}}

\startlongtable
\begin{deluxetable*}{l c c}
\tablewidth{\textwidth}
\tablecolumns{3}
\tablecaption{\swift\ XRT observations\label{tab:xmm}}
\tablehead{MJD & Intergration time 	& Flux\\
& (s) & ($10^{-14}$erg\,s$^{-1}$\,cm$^{-2}$)}
\startdata
58432.20   &   2420    &  $<$6.44  \\
58434.68   &   2220    &  $<$6.32  \\
58436.32   &   2160    &  $<$4.46  \\
58439.58   &   2068    &  $<$5.97  \\
58440.44   &   2562    &  $<$7.26  \\
58442.56   &   2065    &  $<$8.30  \\
58443.56   &   787     &  $<$9.61  \\
58445.48   &   1069    &  $<$7.06  \\
58446.41   &   2530    &  $<$6.77  \\
58449.27   &   1788    &  $<$5.39  \\
58453.85   &   2293    &  $<$6.11  \\
58455.70   &   2595    &  6.61$\pm$3.18  \\
58458.49   &   2415    &  6.95$\pm$3.14  \\
58461.76   &   782     &  $<$9.65  \\
58464.33   &   2080    &  3.95$\pm$2.58  \\
58467.53   &   2417    &  $<$7.39  \\
58475.76   &   2475    &  $<$7.22  \\
58481.29   &   2467    &  $<$7.53  \\
58484.67   &   2325    &  $<$6.03  \\
58487.66   &   2275    &  $<$5.81  \\
58490.44   &   2312    &  $<$5.71  \\
58493.43   &   2372    &  $<$8.14  \\
58496.62   &   2444    &  $<$6.70  \\
58499.41   &   2407    &  9.67$\pm$3.88  \\
58502.00   &   1821    &  $<$5.28  \\
58513.10   &   1563    &  $<$6.77  \\
58518.53   &   2010    &  $<$5.71  \\
58521.46   &   1151    &  $<$9.19  \\
58525.23   &   2210    &  $<$7.41  \\
58536.54   &   2183    &  $<$7.50  \\
58541.05   &   1573    &  $<$7.86  \\
58545.49   &   2462    &  $<$6.02  \\
58549.82   &   2480    &  $<$5.33  \\
58579.09   &   1169    &  $<$9.06  \\
58581.56   &   2467    &  $<$9.22  \\
58584.22   &   1416    &  $<$8.73  \\
58666.09   &   2140    &  $<$4.95  \\
\enddata
\tablecomments{XRT Flux in the 0.3--10keV engergy band is calculated from count rate by assuming an absorbed power-law model with a photon index of $\Gamma=2.7$ and a Galaxtic \ion{H}{1} column density of $2.59\times10^{20}$ cm$^{-2}$.}
\end{deluxetable*}

\startlongtable
\begin{deluxetable}{lccc}
\tablewidth{0pt}
\tablecolumns{4}
\tablecaption{Photometric data of AT~2018hyz \label{tab: phot}}
\tabletypesize{\footnotesize}
\tablehead{\colhead{MJD} & \colhead{Magnitude} & \colhead{Filter} & \colhead{Telescope}}
\startdata
58432.207    & 16.30$\pm$0.04    & UVW2    & \textsl{Swift} \\
58434.720    & 16.40$\pm$0.04    & UVW2    & \textsl{Swift} \\
58436.326    & 16.49$\pm$0.04    & UVW2    & \textsl{Swift} \\
58439.580    & 16.55$\pm$0.04    & UVW2    & \textsl{Swift} \\
58440.439    & 16.61$\pm$0.04    & UVW2    & \textsl{Swift} \\
58442.568    & 16.64$\pm$0.04    & UVW2    & \textsl{Swift} \\
58443.561    & 16.59$\pm$0.05    & UVW2    & \textsl{Swift} \\
58445.479    & 16.65$\pm$0.05    & UVW2    & \textsl{Swift} \\
58446.415    & 16.98$\pm$0.04    & UVW2    & \textsl{Swift} \\
58449.277    & 16.86$\pm$0.04    & UVW2    & \textsl{Swift} \\
58453.853    & 16.92$\pm$0.04    & UVW2    & \textsl{Swift} \\
58455.707    & 17.05$\pm$0.04    & UVW2    & \textsl{Swift} \\
58458.497    & 17.04$\pm$0.04    & UVW2    & \textsl{Swift} \\
58461.761    & 17.14$\pm$0.05    & UVW2    & \textsl{Swift} \\
58464.337    & 17.25$\pm$0.04    & UVW2    & \textsl{Swift} \\
58467.533    & 17.39$\pm$0.04    & UVW2    & \textsl{Swift} \\
58475.764    & 17.49$\pm$0.05    & UVW2    & \textsl{Swift} \\
58481.291    & 17.57$\pm$0.05    & UVW2    & \textsl{Swift} \\
58484.673    & 17.65$\pm$0.05    & UVW2    & \textsl{Swift} \\
58487.662    & 17.69$\pm$0.05    & UVW2    & \textsl{Swift} \\
58490.438    & 17.72$\pm$0.05    & UVW2    & \textsl{Swift} \\
58493.434    & 17.72$\pm$0.05    & UVW2    & \textsl{Swift} \\
58496.618    & 17.76$\pm$0.05    & UVW2    & \textsl{Swift} \\
58499.414    & 17.77$\pm$0.05    & UVW2    & \textsl{Swift} \\
58502.000    & 17.83$\pm$0.05    & UVW2    & \textsl{Swift} \\
58513.104    & 18.05$\pm$0.06    & UVW2    & \textsl{Swift} \\
58518.534    & 17.89$\pm$0.05    & UVW2    & \textsl{Swift} \\
58521.466    & 17.87$\pm$0.06    & UVW2    & \textsl{Swift} \\
58525.235    & 17.90$\pm$0.05    & UVW2    & \textsl{Swift} \\
58536.539    & 17.97$\pm$0.05    & UVW2    & \textsl{Swift} \\
58541.049    & 18.02$\pm$0.06    & UVW2    & \textsl{Swift} \\
58545.498    & 18.04$\pm$0.05    & UVW2    & \textsl{Swift} \\
58549.819    & 18.06$\pm$0.05    & UVW2    & \textsl{Swift} \\
58579.092    & 18.35$\pm$0.07    & UVW2    & \textsl{Swift} \\
58581.563    & 18.51$\pm$0.06    & UVW2    & \textsl{Swift} \\
58584.217    & 18.50$\pm$0.07    & UVW2    & \textsl{Swift} \\
58666.093    & 18.97$\pm$0.07    & UVW2    & \textsl{Swift} \\
58672.068    & 19.29$\pm$0.07    & UVW2    & \textsl{Swift} \\
58432.212    & 15.99$\pm$0.04    & UVM2    & \textsl{Swift} \\
58434.725    & 16.10$\pm$0.04    & UVM2    & \textsl{Swift} \\
58436.331    & 16.15$\pm$0.04    & UVM2    & \textsl{Swift} \\
58439.585    & 16.24$\pm$0.04    & UVM2    & \textsl{Swift} \\
58440.444    & 16.27$\pm$0.04    & UVM2    & \textsl{Swift} \\
58442.573    & 16.31$\pm$0.04    & UVM2    & \textsl{Swift} \\
58443.564    & 16.38$\pm$0.05    & UVM2    & \textsl{Swift} \\
58445.483    & 16.41$\pm$0.04    & UVM2    & \textsl{Swift} \\
58446.419    & 16.61$\pm$0.04    & UVM2    & \textsl{Swift} \\
58449.281    & 16.56$\pm$0.04    & UVM2    & \textsl{Swift} \\
58453.857    & 16.71$\pm$0.04    & UVM2    & \textsl{Swift} \\
58455.712    & 16.81$\pm$0.04    & UVM2    & \textsl{Swift} \\
58458.502    & 16.87$\pm$0.04    & UVM2    & \textsl{Swift} \\
58461.764    & 16.97$\pm$0.05    & UVM2    & \textsl{Swift} \\
58464.340    & 17.06$\pm$0.04    & UVM2    & \textsl{Swift} \\
58467.537    & 17.18$\pm$0.04    & UVM2    & \textsl{Swift} \\
58475.770    & 17.41$\pm$0.05    & UVM2    & \textsl{Swift} \\
58481.294    & 17.57$\pm$0.05    & UVM2    & \textsl{Swift} \\
58484.679    & 17.56$\pm$0.05    & UVM2    & \textsl{Swift} \\
58487.666    & 17.65$\pm$0.05    & UVM2    & \textsl{Swift} \\
58490.441    & 17.61$\pm$0.05    & UVM2    & \textsl{Swift} \\
58493.437    & 17.70$\pm$0.05    & UVM2    & \textsl{Swift} \\
58496.622    & 17.71$\pm$0.05    & UVM2    & \textsl{Swift} \\
58499.418    & 17.70$\pm$0.05    & UVM2    & \textsl{Swift} \\
58502.004    & 17.79$\pm$0.05    & UVM2    & \textsl{Swift} \\
58513.625    & 17.99$\pm$0.06    & UVM2    & \textsl{Swift} \\
58518.538    & 17.83$\pm$0.05    & UVM2    & \textsl{Swift} \\
58521.468    & 17.91$\pm$0.06    & UVM2    & \textsl{Swift} \\
58525.238    & 17.87$\pm$0.05    & UVM2    & \textsl{Swift} \\
58536.542    & 18.03$\pm$0.05    & UVM2    & \textsl{Swift} \\
58541.056    & 17.98$\pm$0.06    & UVM2    & \textsl{Swift} \\
58545.505    & 18.17$\pm$0.05    & UVM2    & \textsl{Swift} \\
58549.823    & 18.15$\pm$0.05    & UVM2    & \textsl{Swift} \\
58579.095    & 18.44$\pm$0.08    & UVM2    & \textsl{Swift} \\
58581.567    & 18.60$\pm$0.06    & UVM2    & \textsl{Swift} \\
58584.219    & 18.67$\pm$0.07    & UVM2    & \textsl{Swift} \\
58666.097    & 19.23$\pm$0.08    & UVM2    & \textsl{Swift} \\
58672.072    & 19.45$\pm$0.08    & UVM2    & \textsl{Swift} \\
58432.203    & 15.99$\pm$0.04    & UVW1    & \textsl{Swift} \\
58434.676    & 16.02$\pm$0.04    & UVW1    & \textsl{Swift} \\
58436.322    & 16.07$\pm$0.04    & UVW1    & \textsl{Swift} \\
58439.576    & 16.18$\pm$0.04    & UVW1    & \textsl{Swift} \\
58440.435    & 16.23$\pm$0.04    & UVW1    & \textsl{Swift} \\
58442.564    & 16.24$\pm$0.04    & UVW1    & \textsl{Swift} \\
58443.559    & 16.29$\pm$0.05    & UVW1    & \textsl{Swift} \\
58445.476    & 16.35$\pm$0.05    & UVW1    & \textsl{Swift} \\
58446.411    & 16.47$\pm$0.04    & UVW1    & \textsl{Swift} \\
58449.274    & 16.41$\pm$0.04    & UVW1    & \textsl{Swift} \\
58453.850    & 16.53$\pm$0.04    & UVW1    & \textsl{Swift} \\
58455.703    & 16.65$\pm$0.04    & UVW1    & \textsl{Swift} \\
58458.492    & 16.77$\pm$0.04    & UVW1    & \textsl{Swift} \\
58461.758    & 16.82$\pm$0.06    & UVW1    & \textsl{Swift} \\
58464.334    & 16.96$\pm$0.05    & UVW1    & \textsl{Swift} \\
58467.530    & 17.04$\pm$0.05    & UVW1    & \textsl{Swift} \\
58475.758    & 17.24$\pm$0.05    & UVW1    & \textsl{Swift} \\
58481.289    & 17.31$\pm$0.05    & UVW1    & \textsl{Swift} \\
58484.669    & 17.46$\pm$0.05    & UVW1    & \textsl{Swift} \\
58487.658    & 17.36$\pm$0.05    & UVW1    & \textsl{Swift} \\
58490.436    & 17.44$\pm$0.05    & UVW1    & \textsl{Swift} \\
58493.431    & 17.49$\pm$0.05    & UVW1    & \textsl{Swift} \\
58496.616    & 17.51$\pm$0.05    & UVW1    & \textsl{Swift} \\
58499.410    & 17.52$\pm$0.05    & UVW1    & \textsl{Swift} \\
58501.997    & 17.53$\pm$0.06    & UVW1    & \textsl{Swift} \\
58513.102    & 17.78$\pm$0.06    & UVW1    & \textsl{Swift} \\
58518.531    & 17.66$\pm$0.06    & UVW1    & \textsl{Swift} \\
58521.464    & 17.77$\pm$0.07    & UVW1    & \textsl{Swift} \\
58525.232    & 17.71$\pm$0.06    & UVW1    & \textsl{Swift} \\
58536.537    & 17.80$\pm$0.06    & UVW1    & \textsl{Swift} \\
58541.044    & 17.86$\pm$0.06    & UVW1    & \textsl{Swift} \\
58545.493    & 17.99$\pm$0.06    & UVW1    & \textsl{Swift} \\
58549.815    & 18.03$\pm$0.06    & UVW1    & \textsl{Swift} \\
58579.090    & 18.33$\pm$0.08    & UVW1    & \textsl{Swift} \\
58581.560    & 18.51$\pm$0.07    & UVW1    & \textsl{Swift} \\
58584.215    & 18.47$\pm$0.08    & UVW1    & \textsl{Swift} \\
58666.090    & 18.84$\pm$0.09    & UVW1    & \textsl{Swift} \\
58672.064    & 19.11$\pm$0.10    & UVW1    & \textsl{Swift} \\
58432.205    & 16.08$\pm$0.04    & U       & \textsl{Swift} \\
58434.678    & 15.96$\pm$0.04    & U       & \textsl{Swift} \\
58436.325    & 16.13$\pm$0.04    & U       & \textsl{Swift} \\
58439.578    & 16.16$\pm$0.04    & U       & \textsl{Swift} \\
58440.437    & 16.17$\pm$0.04    & U       & \textsl{Swift} \\
58442.566    & 16.19$\pm$0.04    & U       & \textsl{Swift} \\
58443.560    & 16.16$\pm$0.06    & U       & \textsl{Swift} \\
58445.478    & 16.22$\pm$0.05    & U       & \textsl{Swift} \\
58446.413    & 16.30$\pm$0.04    & U       & \textsl{Swift} \\
58449.276    & 16.31$\pm$0.04    & U       & \textsl{Swift} \\
58453.852    & 16.45$\pm$0.04    & U       & \textsl{Swift} \\
58455.705    & 16.50$\pm$0.04    & U       & \textsl{Swift} \\
58458.495    & 16.57$\pm$0.04    & U       & \textsl{Swift} \\
58461.760    & 16.66$\pm$0.07    & U       & \textsl{Swift} \\
58464.336    & 16.72$\pm$0.05    & U       & \textsl{Swift} \\
58467.532    & 16.84$\pm$0.05    & U       & \textsl{Swift} \\
58475.762    & 16.97$\pm$0.05    & U       & \textsl{Swift} \\
58481.290    & 17.04$\pm$0.05    & U       & \textsl{Swift} \\
58484.672    & 17.17$\pm$0.05    & U       & \textsl{Swift} \\
58487.660    & 17.27$\pm$0.05    & U       & \textsl{Swift} \\
58490.437    & 17.24$\pm$0.05    & U       & \textsl{Swift} \\
58493.432    & 17.32$\pm$0.05    & U       & \textsl{Swift} \\
58496.617    & 17.35$\pm$0.05    & U       & \textsl{Swift} \\
58499.412    & 17.32$\pm$0.05    & U       & \textsl{Swift} \\
58501.999    & 17.43$\pm$0.06    & U       & \textsl{Swift} \\
58513.104    & 17.56$\pm$0.07    & U       & \textsl{Swift} \\
58518.533    & 17.42$\pm$0.06    & U       & \textsl{Swift} \\
58521.465    & 17.54$\pm$0.08    & U       & \textsl{Swift} \\
58525.234    & 17.57$\pm$0.06    & U       & \textsl{Swift} \\
58536.538    & 17.69$\pm$0.06    & U       & \textsl{Swift} \\
58541.047    & 17.91$\pm$0.08    & U       & \textsl{Swift} \\
58545.496    & 17.77$\pm$0.06    & U       & \textsl{Swift} \\
58549.817    & 17.88$\pm$0.07    & U       & \textsl{Swift} \\
58579.091    & 18.23$\pm$0.10    & U       & \textsl{Swift} \\
58581.562    & 18.34$\pm$0.08    & U       & \textsl{Swift} \\
58584.216    & 18.16$\pm$0.09    & U       & \textsl{Swift} \\
58666.092    & 18.50$\pm$0.11    & U       & \textsl{Swift} \\
58672.067    & 18.59$\pm$0.11    & U       & \textsl{Swift} \\
58438.328    & 16.29$\pm$0.01    & g    & Swope \\
58440.331    & 16.25$\pm$0.01    & g    & Swope \\
58442.343    & 16.29$\pm$0.01    & g    & Swope \\
58443.325    & 16.32$\pm$0.01    & g    & Swope \\
58449.331    & 16.45$\pm$0.01    & g    & Swope \\
58451.335    & 16.54$\pm$0.01    & g    & Swope \\
58460.311    & 16.68$\pm$0.01    & g    & Swope \\
58463.319    & 16.69$\pm$0.01    & g    & Swope \\
58466.290    & 16.76$\pm$0.01    & g    & Swope \\
58468.297    & 16.70$\pm$0.01    & g    & Swope \\
58471.275    & 16.91$\pm$0.01    & g    & Swope \\
58474.345    & 16.95$\pm$0.01    & g    & Swope \\
58480.306    & 17.12$\pm$0.01    & g    & Swope \\
58484.357    & 17.10$\pm$0.01    & g    & Swope \\
58487.336    & 17.10$\pm$0.01    & g    & Swope \\
58491.339    & 17.14$\pm$0.01    & g    & Swope \\
58494.294    & 17.21$\pm$0.02    & g    & Swope \\
58498.203    & 17.29$\pm$0.01    & g    & Swope \\
58511.269    & 17.42$\pm$0.02    & g    & Swope \\
58514.230    & 17.37$\pm$0.01    & g    & Swope \\
58519.346    & 17.46$\pm$0.01    & g    & Swope \\
58521.206    & 17.47$\pm$0.01    & g    & Swope \\
58525.338    & 17.38$\pm$0.01    & g    & Swope \\
58528.392    & 17.68$\pm$0.02    & g    & Swope \\
58540.291    & 17.70$\pm$0.01    & g    & Swope \\
58544.102    & 17.67$\pm$0.01    & g    & Swope \\
58551.224    & 17.65$\pm$0.01    & g    & Swope \\
58556.136    & 17.89$\pm$0.01    & g    & Swope \\
58559.232    & 17.88$\pm$0.01    & g    & Swope \\
58571.080    & 17.87$\pm$0.01    & g    & Swope \\
58579.176    & 17.93$\pm$0.01    & g    & Swope \\
58585.122    & 17.98$\pm$0.01    & g    & Swope \\
58588.069    & 18.02$\pm$0.01    & g    & Swope \\
58600.036    & 18.14$\pm$0.02    & g    & Swope \\
58609.010    & 18.40$\pm$0.01    & g    & Swope \\
58616.073    & 18.42$\pm$0.01    & g    & Swope \\
58637.040    & 18.42$\pm$0.01    & g    & Swope \\
58643.007    & 18.16$\pm$0.01    & g    & Swope \\
58659.976    & 18.20$\pm$0.01    & g    & Swope \\
58674.970    & 18.57$\pm$0.04    & g    & Swope \\
58438.321    & 16.56$\pm$0.00    & r    & Swope \\
58440.323    & 16.49$\pm$0.00    & r    & Swope \\
58442.350    & 16.56$\pm$0.00    & r    & Swope \\
58443.317    & 16.53$\pm$0.01    & r    & Swope \\
58447.321    & 16.61$\pm$0.01    & r    & Swope \\
58449.339    & 16.63$\pm$0.01    & r    & Swope \\
58451.328    & 16.68$\pm$0.01    & r    & Swope \\
58460.319    & 16.90$\pm$0.01    & r    & Swope \\
58463.312    & 16.91$\pm$0.01    & r    & Swope \\
58466.297    & 16.98$\pm$0.01    & r    & Swope \\
58468.291    & 17.02$\pm$0.01    & r    & Swope \\
58471.282    & 17.10$\pm$0.01    & r    & Swope \\
58474.339    & 17.16$\pm$0.01    & r    & Swope \\
58480.299    & 17.37$\pm$0.01    & r    & Swope \\
58484.352    & 17.38$\pm$0.01    & r    & Swope \\
58487.329    & 17.44$\pm$0.01    & r    & Swope \\
58491.332    & 17.49$\pm$0.01    & r    & Swope \\
58494.287    & 17.56$\pm$0.01    & r    & Swope \\
58498.195    & 17.66$\pm$0.01    & r    & Swope \\
58511.262    & 17.85$\pm$0.01    & r    & Swope \\
58514.223    & 17.88$\pm$0.01    & r    & Swope \\
58519.354    & 18.08$\pm$0.01    & r    & Swope \\
58521.194    & 17.94$\pm$0.01    & r    & Swope \\
58525.349    & 18.05$\pm$0.02    & r    & Swope \\
58527.380    & 18.13$\pm$0.01    & r    & Swope \\
58540.280    & 18.34$\pm$0.01    & r    & Swope \\
58544.113    & 18.29$\pm$0.01    & r    & Swope \\
58551.204    & 18.45$\pm$0.01    & r    & Swope \\
58556.122    & 18.59$\pm$0.01    & r    & Swope \\
58559.216    & 18.71$\pm$0.01    & r    & Swope \\
58571.065    & 18.86$\pm$0.01    & r    & Swope \\
58579.161    & 18.98$\pm$0.01    & r    & Swope \\
58585.139    & 19.08$\pm$0.01    & r    & Swope \\
58588.054    & 18.95$\pm$0.02    & r    & Swope \\
58600.020    & 19.28$\pm$0.02    & r    & Swope \\
58608.994    & 19.39$\pm$0.08    & r    & Swope \\
58608.995    & 19.33$\pm$0.02    & r    & Swope \\
58616.057    & 19.59$\pm$0.03    & r    & Swope \\
58637.025    & 19.61$\pm$0.02    & r    & Swope \\
58642.984    & 19.68$\pm$0.04    & r    & Swope \\
58642.992    & 19.37$\pm$0.02    & r    & Swope \\
58659.961    & 19.60$\pm$0.03    & r    & Swope \\
58438.324    & 16.63$\pm$0.00    & i    & Swope \\
58440.327    & 16.56$\pm$0.00    & i    & Swope \\
58442.347    & 16.58$\pm$0.00    & i    & Swope \\
58443.321    & 16.58$\pm$0.01    & i    & Swope \\
58447.326    & 16.64$\pm$0.01    & i    & Swope \\
58449.335    & 16.63$\pm$0.00    & i    & Swope \\
58451.331    & 16.71$\pm$0.01    & i    & Swope \\
58460.315    & 16.87$\pm$0.01    & i    & Swope \\
58463.315    & 16.88$\pm$0.00    & i    & Swope \\
58466.294    & 16.96$\pm$0.01    & i    & Swope \\
58468.294    & 17.02$\pm$0.01    & i    & Swope \\
58471.279    & 17.07$\pm$0.01    & i    & Swope \\
58474.342    & 17.07$\pm$0.01    & i    & Swope \\
58480.302    & 17.33$\pm$0.01    & i    & Swope \\
58487.332    & 17.41$\pm$0.01    & i    & Swope \\
58491.336    & 17.44$\pm$0.01    & i    & Swope \\
58494.290    & 17.50$\pm$0.01    & i    & Swope \\
58498.199    & 17.63$\pm$0.01    & i    & Swope \\
58511.266    & 17.83$\pm$0.01    & i    & Swope \\
58514.226    & 17.85$\pm$0.01    & i    & Swope \\
58519.350    & 18.07$\pm$0.01    & i    & Swope \\
58521.200    & 17.98$\pm$0.01    & i    & Swope \\
58525.344    & 18.18$\pm$0.02    & i    & Swope \\
58527.386    & 18.18$\pm$0.03    & i    & Swope \\
58540.286    & 18.34$\pm$0.01    & i    & Swope \\
58551.211    & 18.48$\pm$0.01    & i    & Swope \\
58551.218    & 18.48$\pm$0.01    & i    & Swope \\
58559.224    & 18.74$\pm$0.02    & i    & Swope \\
58571.073    & 18.80$\pm$0.01    & i    & Swope \\
58579.169    & 18.87$\pm$0.02    & i    & Swope \\
58585.131    & 19.08$\pm$0.02    & i    & Swope \\
58588.062    & 18.91$\pm$0.02    & i    & Swope \\
58600.028    & 19.28$\pm$0.02    & i    & Swope \\
58616.065    & 19.50$\pm$0.03    & i    & Swope \\
58637.033    & 19.58$\pm$0.03    & i    & Swope \\
58659.968    & 19.79$\pm$0.04    & i    & Swope \\

\enddata
\tablecomments{All the data presented in this table have been Galactic extinction corrected. We did not perform host subtraction on the \swift\ UV data since the host contribution is negligible. On the other hand, the Swope data are host-subtracted.}
\end{deluxetable}

\end{document}